%Paper: solv-int/9401003
%From: clarkson@maths.ex.ac.uk
%Date: Thu, 13 Jan 94 16:13:33 GMT

\magnification=\magstephalf
\baselineskip=13.5pt

\hsize=6.5 truein
\vsize=9.5 truein
\hfuzz=2pt\vfuzz=4pt
\pretolerance=5000
\tolerance=5000
\parskip=0pt plus 1pt
\parindent=16pt
\font\fourteenrm=cmr10 scaled \magstep2
\font\fourteeni=cmmi10 scaled \magstep2
\font\fourteenbf=cmbx10 scaled \magstep2
\font\fourteenit=cmti10 scaled \magstep2
\font\fourteensy=cmsy10 scaled \magstep2
\font\large=cmbx10 scaled \magstep1

\font\eightrm=cmr8
\font\eighti=cmmi8
\font\eightbf=cmbx8
\font\eightit=cmti8

\font\eightsy=cmsy8
\font\sixrm=cmr6
\font\sixi=cmmi6
\font\sixsy=cmsy6

\def\tenpoint{\def\rm{\fam0\tenrm}%
  \textfont0=\tenrm \scriptfont0=\sevenrm
		      \scriptscriptfont0=\fiverm
  \textfont1=\teni  \scriptfont1=\seveni
		      \scriptscriptfont1=\fivei
  \textfont2=\tensy \scriptfont2=\sevensy
		      \scriptscriptfont2=\fivesy
  \textfont3=\tenex   \scriptfont3=\tenex
		      \scriptscriptfont3=\tenex
  \textfont\itfam=\tenit  \def\it{\fam\itfam\tenit}%
  \textfont\slfam=\tensl  \def\sl{\fam\slfam\tensl}%
  \textfont\bffam=\tenbf  \scriptfont\bffam=\sevenbf
			    \scriptscriptfont\bffam=\fivebf
			    \def\bf{\fam\bffam\tenbf}%
  \normalbaselineskip=20 truept
  \setbox\strutbox=\hbox{\vrule height14pt depth6pt width0pt}%
  \let\sc=\eightrm \normalbaselines\rm}
\def\eightpoint{\def\rm{\fam0\eightrm}%
  \textfont0=\eightrm \scriptfont0=\sixrm
		      \scriptscriptfont0=\fiverm
  \textfont1=\eighti  \scriptfont1=\sixi
		      \scriptscriptfont1=\fivei
  \textfont2=\eightsy \scriptfont2=\sixsy
		      \scriptscriptfont2=\fivesy
  \textfont3=\tenex   \scriptfont3=\tenex
		      \scriptscriptfont3=\tenex
  \textfont\itfam=\eightit  \def\it{\fam\itfam\eightit}%
  \textfont\bffam=\eightbf  \def\bf{\fam\bffam\eightbf}%
  \normalbaselineskip=16 truept
  \setbox\strutbox=\hbox{\vrule height11pt depth5pt width0pt}}
\def\fourteenpoint{\def\rm{\fam0\fourteenrm}%
  \textfont0=\fourteenrm \scriptfont0=\tenrm
		      \scriptscriptfont0=\eightrm
  \textfont1=\fourteeni  \scriptfont1=\teni
		      \scriptscriptfont1=\eighti
  \textfont2=\fourteensy \scriptfont2=\tensy
		      \scriptscriptfont2=\eightsy
  \textfont3=\tenex   \scriptfont3=\tenex
		      \scriptscriptfont3=\tenex
  \textfont\itfam=\fourteenit  \def\it{\fam\itfam\fourteenit}%
  \textfont\bffam=\fourteenbf  \scriptfont\bffam=\tenbf
			     \scriptscriptfont\bffam=\eightbf
			     \def\bf{\fam\bffam\fourteenbf}%
  \normalbaselineskip=24 truept
  \setbox\strutbox=\hbox{\vrule height17pt depth7pt width0pt}%
  \let\sc=\tenrm \normalbaselines\rm}

\def\today{\number\day\ \ifcase\month\or
  January\or February\or March\or April\or May\or June\or
  July\or August\or September\or October\or November\or December\fi
  \space \number\year}
\newcount\secno      %section number
\newcount\subno      %number of subsection
\newcount\subsubno   %number of subsubsection
\newcount\appno      %appendix number
\newcount\tableno    %table number
\newcount\figureno   %figure number
\normalbaselineskip=15 truept
\baselineskip=15 truept
\def\title#1
   {\vglue1truein
   {\baselineskip=24 truept
    \pretolerance=10000
    \raggedright
    \noindent \fourteenpoint\bf #1\par}
    \vskip1truein minus36pt}
\def\author#1
  {{\pretolerance=10000
    \raggedright
    \noindent {\large #1}\par}}
\def\address#1
   {\bigskip
    \noindent \rm #1\par}
\def\shorttitle#1
   {\vfill
    \noindent \rm Short title: {\sl #1}\par
    \medskip}
\def\pacs#1
   {\noindent \rm PACS number(s): #1\par
    \medskip}
\def\jnl#1
   {\noindent \rm Submitted to: {\sl #1}\par
    \medskip}
\def\date
   {\noindent Date: \today\par
    \medskip}
\def\beginabstract
   {\vskip 1in\baselineskip=12pt
    \noindent {\bf Abstract. }\smallskip\rm}
\def\keyword#1
   {\bigskip
    \noindent {\bf Keyword abstract: }\rm#1}
\def\endabstract
   {\par
    \vfill\eject}

\def\entry#1#2#3
   {\noindent
    \hangindent=20pt
    \hangafter=1
    \hbox to20pt{#1 \hss}#2\hfill #3\par}
\def\subentry#1#2#3
   {\noindent
    \hangindent=40pt
    \hangafter=1
    \hskip20pt\hbox to20pt{#1 \hss}#2\hfill #3\par}
\def\section#1
   {\vskip0pt plus.1\vsize\penalty-250
    \vskip0pt plus-.1\vsize\vskip24pt plus12pt minus6pt
    \subno=0 \subsubno=0
    \global\advance\secno by 1
    \noindent {\bf \the\secno. #1\par}
    \bigskip
    \noindent}
\def\subsection#1
   {\vskip-\lastskip
    \vskip24pt plus12pt minus6pt
    \bigbreak
    \global\advance\subno by 1
    \subsubno=0
    \noindent {\sl \the\secno.\the\subno. #1\par}
    \nobreak
    \medskip
    \noindent}
\def\subsubsection#1
   {\vskip-\lastskip
    \vskip20pt plus6pt minus6pt
    \bigbreak
    \global\advance\subsubno by 1
    \noindent {\sl \the\secno.\the\subno.\the\subsubno. #1}\null. }
\def\appendix#1
   {\vskip0pt plus.1\vsize\penalty-250
    \vskip0pt plus-.1\vsize\vskip24pt plus12pt minus6pt
    \subno=0 \eqnno=0
    \global\advance\appno by 1
    \noindent {\bf Appendix \the\appno. #1\par}
    \bigskip
    \noindent}
\def\subappendix#1
   {\vskip-\lastskip
    \vskip36pt plus12pt minus12pt
    \bigbreak
    \global\advance\subno by 1
    \noindent {\sl \the\appno.\the\subno. #1\par}
    \nobreak
    \medskip
    \noindent}
\def\ack
   {\vskip-\lastskip
    \vskip36pt plus12pt minus12pt
    \bigbreak
    \noindent{\bf Acknowledgements\par}
    \nobreak
    \bigskip
    \noindent}

\def\tabcaption#1
   {\global\advance\tableno by 1
    \noindent {\bf Table \the\tableno.} \rm#1\par
    \bigskip}
\def\figures
   {\vskip 1in %\vfill\eject
    \noindent {\bf Figure captions\par}
    \bigskip}
\def\figcaption#1
   {\global\advance\figureno by 1
    \noindent {\bf Figure \the\figureno.} \rm#1\par
    \bigskip}
\def\references
     {\vskip .5in %\vfill\eject
     {\noindent \bf References\par}
      \parindent=0pt
      \bigskip}
\def\refjl#1#2#3#4
   {\hangindent=16pt
    \hangafter=1
    \rm #1
   {\frenchspacing\sl #2
    \bf #3}
    #4\par}
\def\refbk#1#2#3
   {\hangindent=16pt
    \hangafter=1
    \rm #1
   {\frenchspacing\sl #2}
    #3\par}
\def\numrefjl#1#2#3#4#5
   {\parindent=40pt
    \hang
    \noindent
    \rm {\hbox to 30truept{\hss #1\quad}}#2
   {\frenchspacing\sl #3\/
    \bf #4}
    #5\par\parindent=16pt}
\def\numrefbk#1#2#3#4
   {\parindent=40pt
    \hang
    \noindent
    \rm {\hbox to 30truept{\hss #1\quad}}#2
   {\frenchspacing\sl #3\/}
    #4\par\parindent=16pt}

\def\frac#1#2{{#1 \over #2}}

\def\d{{\rm d}}

\def\i{\ifmmode{\rm i}\else\char"10\fi}
\def\case#1#2{{\textstyle{#1\over #2}}}
\def\boldrule#1{\vbox{\hrule height1pt width#1}}

\def\etal{{\sl et al\/}\ }
\catcode`\@=11
\def\ind{\hbox to 5pc{}}
\def\eq(#1){\hfill\llap{(#1)}}

\def\deqn#1{\displ@y\halign{\hbox to \displaywidth
    {$\@lign\displaystyle##\hfil$}\crcr #1\crcr}}
\def\indeqn#1{\displ@y\halign{\hbox to \displaywidth
    {$\ind\@lign\displaystyle##\hfil$}\crcr #1\crcr}}
\def\indalign#1{\displ@y \tabskip=0pt
  \halign to\displaywidth{\ind$\@lign\displaystyle{##}$\tabskip=0pt
    &$\@lign\displaystyle{{}##}$\hfill\tabskip=\centering
    &\llap{$\@lign##$}\tabskip=0pt\crcr
    #1\crcr}}
\catcode`\@=12

\def\IP{Inverse Problems}
\def\JPA{J. Phys. A: Math. Gen.}
        %1968-87
   %1988 and onwards
     %1968--1988
        %1989 and onwards

           %1975--1988
     %1989 and onwards

\def\JMP{J. Math. Phys.}

\def\JPSJ{J. Phys. Soc. Japan}

\def\PRL{Phys. Rev. Lett.}

\font\sc=cmcsc10

\def\p{Pain\-lev\'e}
\def\d{{\rm d}}\def\i{{\rm i}}

\def\tfr#1#2{{\tx{#1\over#2}}}

\newcount\refno
\def\ref#1#2#3#4#5{\vskip.9pt\global\advance\refno by 1
\item{[{\bf\the\refno}]\ }{\rm#1}, {\it#2}, {\bf#3} (#4) #5}

%\hfill\break}
\def\sam{Stud.\ Appl.\ Math.}
\def\pl{Phys.\ Lett.}
\def\jpa{J.\ Phys.\ A: Math.\ Gen.}
\def\jmp{J.\ Math.\ Phys}

%Clarkson and Kruskal (1989)}
\def\~#1{{\bf\tilde{\mit#1}}}

%{$3+1$-{\smc dzk}}
%{$2+1$-{\smc dzk}}
%{$3+1$-{\smc zk}}

\def\secn{\the\secno}

\font\sit=cmti9

%{{\partial\over\partial#1}}%

\nopagenumbers
\def\sch{Schr\"odinger}

\def\and{\qquad {\rm and}\qquad}

\def\tfr#1#2{{\tx{#1\over#2}}}

\def\pde{partial differential equa\-tion}%{{\sc pde}}%
\def\pdes{partial differential equa\-tions}
\def\ode{ordinary differential equa\-tion}
\def\odes{ordinary differential equa\-tions}

\def\eq{equa\-tion}

\def\dgb{Differ\-ential Gr\"obner Bases}

\def\d{{\rm d}}\def\i{{\rm i}}
\def\yp{{\d y \over\d x}}

\def\wz{{\d w \over\d z}}
\def\wzz{{\d^2w \over\d z^2}}
\def\wzzz{{\d^3w \over\d z^3}}
\def\wzzzz{{\d^4w \over\d z^4}}
\def\vz{{\d v \over\d \zeta}}
\def\vzz{{\d^2v \over\d \zeta^2}}

\def\vzzzz{{\d^4v \over\d \zeta^4}}

\def\sech{\mathop{\rm sech}\nolimits}

%\font\teneuf=eufm10
%\font\seveneuf=eufm scaled 700
%\font\fiveeuf=eufm scaled 500
%\newfam\euffam
%\textfont\euffam=\teneuf \scriptfont\euffam=\seveneuf
%   \scriptscriptfont\euffam=\fiveeuf

%\def\frak#1{{\fam\euffam\relax#1}}

%\def\fg{\frak g} \def\fh{\frak h}

\newcount\exno
\newcount\secno   %number of section
\newcount\subno   %number of subsection
\newcount\subsubno   %number of subsubsection
\newcount\figno
\newcount\tableno

\def\section#1
   {\vskip0pt plus.1\vsize\penalty-250
    \vskip0pt plus-.1\vsize\vskip18pt plus9pt minus6pt
     \subno=0 \exno=0 \figno=0  \eqnno=0
   {\parindent=30pt\raggedright
    \global\advance\secno by 1
    \item{\hbox to 25pt{\large\the\secno\hfill}}\large #1.
    \medskip}}
\def\subsection#1
   {\vskip-\lastskip
    \exno=0 \caseno=0 \tableno=0 \subsubno=0
    \vskip15pt plus4pt minus4pt
    \bigbreak
    \global\advance\subno by 1
    \noindent {\bf \the\secno.\the\subno\enskip #1. }}

\def\subsubsection#1
   {\vskip-\lastskip
    \exno=0 \caseno=0
     \vskip4pt plus2pt minus2pt
    \bigbreak \global\advance\subsubno by 1
   \noindent {\sl \the\secno.\the\subno.\the\subsubno\enskip #1. }}

\def\boldrule#1{\vbox{\hrule height1pt width#1}}

\def\bline#1{\boldrule{#1truein}}

\def\Table#1#2#3{\vskip-\lastskip
    \vskip4pt plus2pt minus2pt
    \bigbreak \global\advance\tableno by 1
		 \vbox{\centerline{\bf Table \the\secno.\the\subno%
		 \uppercase\expandafter{\romannumeral\the\tableno}}\smallskip
		      \centerline{\sl #1}
 {$$\vbox{\offinterlineskip\tabskip=0pt
       \bline{#2}
       \halign to#2truein{#3}
       \bline{#2}}
   $$}}}

\newbox\strutbox
\setbox\strutbox=\hbox{\vrule height10pt depth 4.5pt width0pt}

\def\defn#1{\vskip-\lastskip
    \vskip4pt plus2pt minus2pt% \vskip8pt plus4pt minus4pt
    \bigbreak
    \global\advance\exno by 1
    \noindent {{\bf Definition\ \the\secno.\the\subno.\the\exno}\enskip
{\rm#1\/}}\smallskip}
\def\example#1{\vskip-\lastskip
    \vskip4pt plus2pt minus2pt \caseno=0
    \bigbreak \global\advance\exno by 1
    \noindent {{\bf Example\ \the\secno.\the\subno.\the\exno}
\enskip{\rm#1}}\smallskip}
\def\exercise#1{\vskip-\lastskip
    \vskip4pt plus2pt minus2pt \caseno=0
    \bigbreak \global\advance\exno by 1
    \noindent {{\bf Exercise\ \the\secno.\the\subno.\the\exno}
\enskip{\rm#1}}\smallskip}
\def\thm#1{\vskip-\lastskip
    \vskip4pt plus2pt minus2pt \caseno=0
    \bigbreak \global\advance\exno by 1
    \noindent {{\bf Theorem\ \the\secno.\the\subno.\the\exno}\enskip
    {\sl#1\/}}\smallskip}
\def\lem#1{\vskip-\lastskip
    \vskip4pt plus2pt minus2pt \caseno=0
    \bigbreak \global\advance\exno by 1
    \noindent {{\bf Lemma\ \the\secno.\the\subno.\the\exno}\enskip
    {\it#1\/}}\smallskip}
\def\remark{\vskip-\lastskip
    \vskip4pt plus2pt minus2pt \caseno=0
    \bigbreak \global\advance\exno by 1
    \noindent {{\bf Remark\ \the\secno.\the\subno.\the\exno}\enskip}}
\def\remarks{\vskip-\lastskip
    \vskip4pt plus2pt minus2pt \caseno=0
    \bigbreak \global\advance\exno by 1
    \noindent {{\bf Remarks\ \the\secno.\the\subno.\the\exno}\enskip}}

\def\=#1{{\bf\bar{\mit#1}}}
\def\^#1{{\bf\hat{\mit#1}}}
\def\~#1{{\bf\tilde{\mit#1}}}

\def\cc#1{\kappa_{#1}}
\def\pr#1{\mathop{\rm pr}\nolimits^{(#1)}}
\def\hbb#1#2#3{\qquad{\hbox to 100pt{$#1$\hfill}}{\hbox to
200pt{$#2$\hfill}}\hfill#3}
\def\ra{{\hbox to 100pt{\hfill}}\Rightarrow\qquad}

\newcount\remno
\def\tfr#1#2{{\textstyle{#1\over#2}}}
\def\rem#1{\global\advance\remno by 1
\item{\the\remno.\enskip}#1}

\newcount\eqnno
\def\sen{\the\secno}%{\the\secno.\the\subno}
\def\eqn#1{\global\advance\eqnno by 1
	   \eqno(\sen.\the\eqnno)
	   \expandafter \xdef\csname #1\endcsname
	   {\sen.\the\eqnno}\relax }
\def\eqnn#1{\global\advance\eqnno by 1
	   (\sen.\the\eqnno)
	   \expandafter \xdef\csname #1\endcsname
	   {\sen.\the\eqnno}\relax }
\def\eqnm#1#2{\global\advance\eqnno by 1
	   (\sen.\the\eqnno\hbox{#2})
	   \expandafter \xdef\csname #1\endcsname
	   {\sen.\the\eqnno}\relax }
\def\eqnr#1{(\sen.\the\eqnno\hbox{#1})}

\newcount\caseno
\def\case#1{\vskip-\lastskip
    \vskip4pt plus2pt minus2pt
    \bigbreak \global\advance\caseno by 1
\noindent\underbar{{\sc Case}
\the\secno.\the\subno.\the\caseno}\enskip{#1}. }

\def\Case#1{\vskip-\lastskip
    \vskip4pt plus2pt minus2pt
    \bigbreak \global\advance\caseno by 1
\noindent {\sc
Case}\ \the\secno.\the\subno.\the\exno{\romannumeral\the\caseno}\enskip
\underbar{#1}.\quad}

\def\paper#1#2#3#4#5#6
   {\par
    \hangindent=16pt\hangafter=1
    {\sc #1}\ {\rm[#2], #6},
   {\frenchspacing\it #3
    \bf #4},    #5.\par}

\def\refpp#1#2#3{}{}{}
\def\refjl#1#2#3#4#5{}{}{}{}{}
\def\refbk#1#2#3#4{}{}{}{}
\def\refcf#1#2#3#4#5#6{}{}{}{}{}{}
\newcount\refno
\refno=0
\def\refn#1{\global\advance\refno by 1
\expandafter \xdef\csname #1\endcsname {\the\refno}\relax}
\def\hide#1{}

\refn{refWhitham}
\refbk{Whitham G B}{1974}{Linear and Nonlinear Waves}{Wiley, New York}

\refn{refHS}
\refjl{Hirota R and Satsuma J}{1976}{J. Phys. Soc. Japan}{40}{611--612}
\refn{refAKNS}
\refjl{Ablowitz M J, Kaup D J, Newell A C and Segur H}{1974}{Stud.
Appl. Math.}{53}{249--315}
\refn{refGGKM}
\refjl{Gardner C S, Greene J M, Kruskal M D and Miura R M}{1967}{Phys.
Rev. Lett}{19}{1095--1097}
\refn{refPer}
\refjl{Peregrine H}{1966}{J. Fluid Mech.}{25}{321--330}
\refn{refBBM}
\refjl{Benjamin T B, Bona J L and Mahoney J}{1972}{Phil. Trans. R. Soc.
Lond. Ser. A}{272}{47--78}
\refn{refMcLO}
\refjl{McLeod J B and Olver P J}{1983}{SIAM
J.\ Math.\ Anal.}{14}{488--506}

\refn{refHiet}
\refcf{Hietarinta J}{1990}{Partially Integrable Evolution Equations in
Physics}{Eds.\ R.\ Conte and
N.\ Boccara}{{\it NATO ASI Series C: Mathematical and Physical
Sciences\/}, {\bf 310}, Kluwer,
Dordrecht}{pp459--478}
\refn{refHirt}
\refcf{Hirota R}{1980}{Solitons}{Eds.\ R.K.\ Bullough and
P.J.\ Caudrey}{{\it Topics in Current
Physics\/}, {\bf 17}, Springer-Verlag, Berlin}{pp157--176}

\refn{refARSa}
\refjl{Ablowitz M J, Ramani A and Segur H}{1978}{\PRL}{23}{333--338}
\refn{refARSb}
\refjl{Ablowitz M J, Ramani A and Segur H}{1980}{\jmp}{21}{715--721}
\refn{refWTC}
\refjl{Weiss J, Tabor M and Carnevale G}{1983}{\jmp}{24}{522--526}

\refn{refBLMP}
\refjl{Boiti M, Leon J J-P, Manna M and Pempinelli
F}{1986}{\IP}{2}{271--279}
\refn{refJM}
\refjl{Jimbo M and Miwa T}{1983}{Publ. R.I.M.S.}{19}{943--1001}
\refn{refDGRW}
\refjl{Dorizzi B, Grammaticos B, Ramani A and Winternitz
P}{1986}{\JMP}{27}{2848--2852}

\refn{refYOS}
\refjl{Yajima N, Oikawa M and Satsuma J}{1978}{\JPSJ}{44}{1711--1714}
\refn{refKY}
\refjl{Kako F and Yajima N}{1980}{\JPSJ}{49}{2063--2071}

\refn{refBogi}
\refjl{Bogoyaviemskii O I}{1990}{Math. USSR Izves.}{34}{245--259}
\refn{refBogii}
\refjl{Bogoyaviemskii O I}{1990}{Russ. Math. Surv.}{45}{1--86}

\refn{refBK}
\refbk{Bluman G W and Kumei S}{1989}{Symmetries and Differential
Equations}{{\fit Appl. Math. Sci.\/}, {\bf 81}, Springer-Verlag,
Berlin}
\refn{refOlver}
\refbk{Olver P J}{1993}{Applications of Lie Groups to Differential
Equations}{2nd Edition, Springer Verlag, New York}

\refn{refHere}
\refjl{Hereman W}{1993}{Euromath Bull.}{2}{to appear}
\refn{refCHW}
\refjl{Champagne B, Hereman W and Winternitz P}{1991}{Comp. Phys.
Comm.}{66}{319--340}

\refn{refBCa}
\refjl{Bluman G W and Cole J D}{1969}{J.\ Math.\ Mech.}{18}{1025--1042}

\refn{refLW}
 \refjl{Levi D and Winternitz P}{1989}{\jpa}{22}{2915--2924}

\refn{refVor}
\refjl{Vorob'ev E M}{1991}{Acta Appl. Math.}{24}{1--24}

 \refn{refCK}
\refjl{Clarkson P A and Kruskal M D}{1989}{\jmp}{30}{2201--2213}

 \refn{refPACrev}
 \refjl{Clarkson P A}{1993}{Math. Comp. Model.}{18}{45--68}
 \refn{refFush}
\refjl{Fushchich W I}{1991}{Ukrain. Mat. Zh.}{43}{1456--1470}

 \refn{refGalaka}
\refjl{Galaktionov V A}{1990}{Diff. and Int. Eqns.}{3}{863--874}
 \refn{refGDEKS}
 \refjl{Galaktionov V A, Dorodnytzin V A, Elenin G G, Kurdjumov S P and
 Samarskii A A}{1988}{J. Sov. Math.}{41}{1222--1292}
 \refn{refAmesii}
 \refjl{Ames W F}{1992}{Appl. Num. Math.}{10}{235--259}
 \refn{refShok}
 \refbk{Shokin Yu I}{1983}{The Method of Differential
 Approximation}{Springer-Verlag, New
York}

 \refn{refReida}
\refjl{Reid G J}{1990}{\jpa}{23}{L853--L859}
 \refn{refReidb}
\refjl{Reid G J}{1991}{Europ. J. Appl. Math.}{2}{293--318}
\refn{refMF}
\refpp{Mansfield E and Fackerell E}{1992}{``Differential Gr\"obner
Bases",
preprint {\bf 92/108}, Macquarie University, Sydney, Australia}
\refn{refCMa}
\refjl{Clarkson P A and Mansfield E L}{1994}{Physica D}{70}{250--288}
\refn{refZwil}
\refbk{Zwillinger D}{1992}{Handbook of Differential Equations}{Second
Edition, Academic,  Boston}

 \refn{refCMb}
\refjl{Clarkson P A and Mansfield E L}{1994}{SIAM J. Appl. Math.}{}{to
appear}
\refn{refSchw}
\refjl{Schwarz F}{1992}{Computing}{49}{95--115}
\refn{refTop}
\refjl{Topunov V L}{1989}{Acta Appl. Math.}{16}{191--206}
\refn{refBuchi}
\refcf{Buchberger B}{1988}{Mathematical Aspects of Scientific
Software}{\rm Ed.\ J.\ Rice}{Springer
Verlag}{pp59--87}

\refn{refPank}
\refjl{Pankrat'ev E V}{1989}{Acta Appl. Math.}{16}{167--189}

\refn{refRW}
\refpp{Reid G J and Wittkopf A}{1993}{``A Differential Algebra Package
for
{\sc maple}'', {\tt ftp 137.82.36.21} login: anonymous, password: your
email address,
directory: {\tt pub/standardform}}

\refn{refMD}
\refpp{Mansfield E}{1993}{``{\tt diffgrob2}: A symbolic algebra package
for analysing systems of PDE
using Maple", {\tt ftp 137.111.216.12},  login: anonymous, password:
your email address,
directory: {\tt pub/maths/Maple}, files:{\tt diffgrob2\_src.tar.Z,
diffgrob2\_man.tex.Z}}

\refn{refWeiss}
\refjl{Weiss J}{1983}{\jmp}{24}{1405--1413}

\refn{refInce}
\refbk{Ince E L}{1956}{Ordinary Differential Equations}{Dover, New
York}
\refn{refWW}
\refbk{Whittaker E E and  Watson G M}{1927}{Modern Analysis}{4th
Edition, C.U.P., Cambridge}
 \refn{refAndIb}
\refbk{Anderson R L and Ibragimov N H}{1979}{Lie-B\"acklund
Transformations in
Applications}{SIAM, Philadelphia}
\refn{refNTZ}
\refjl{Newell A C, Tabor M and Zeng Y B}{1987}{Physica}{29D}{1--68}

\refn{refJWx}
\refcf{Weiss J}{1990}{Solitons in Physics, Mathematics and Nonlinear
Optics}{Eds P.J.\ Olver and
D.H.\ Sattinger}{{\it IMA Series\/}, {\bf 25}, Springer-Verlag,
Berlin}{pp175--202}

\refn{refMLD}
\refjl{Musette M, Lambert F and Decuyper J
C}{1987}{\JPA}{20}{6223--6235}
\refn{refHI}
\refjl{Hirota R and Ito M}{1983}{\JPSJ}{52}{744--748}

\refn{refAC}
\refbk{Ablowitz M J and Clarkson P A}{1991}{Solitons, Nonlinear
Evolution Equations and Inverse
 Scattering}{{\frenchspacing\it L.M.S. Lect. Notes Math.}, {\bf 149},
 C.U.P., Cambridge}
\refn{refCM}
\refjl{Conte R and Musette M}{1991}{\jmp}{32}{1450--1457}
\refn{refHietb}
\refjl{Hietarinta J}{1987}{\jmp}{28}{1732--1742}
\refn{refHL}
\refjl{Hu X B and Li Y}{1983}{\JPA}{24}{1979--1986}
\refn{refMat}
\refjl{Matsuno Y}{1990}{\JPSJ}{59}{3093--3100}
\refn{refMus}
\refcf{Musette M}{1987}{\p\ Transcendents: Their Asymptotics and
Physical Applications}{Eds.\ D.\
Levi and P.\ Winternitz}{{\it NATO ASI Series B: Physics\/}, {\bf
278},  Plenum, New
York}{pp197--209}
\refn{refTag}
\refjl{Tagami Y}{1989}{\pl}{141A}{116--120}
\refn{refJWv}
\refjl{Weiss J}{1985}{\JMP}{26}{2174--2180}

\refn{refDTT}
\refjl{Deift P, Tomei C and Trubowitz E}{1982}{Commun. Pure Appl.
Math.}{35}{567--628}
\refn{refCole}\refjl{Cole J D}{1951}{Quart.\ Appl.\ Math.}{9}{225--236}
\refn{refHopf}\refjl{Hopf E}{1950}{Commun.\ Pure
Appl.\ Math.}{3}{201--250}
\refn{refASH}\refjl{Ablowitz M J, Schober C and Herbst B
M}{1993}{\prl}{71}{2683--2686}
\refn{refOlverb}
\refpp{Olver P J}{1993}{``Direct reduction and differential
constraints", preprint, Department of
Mathematics, University of Maryland, College Park, MD}
\refn{refABH}
\refjl{Arrigo D J, Broadbridge P and Hill J
M}{1993}{\jmp}{34}{4692--4703}
\refn{refPucci}
\refjl{Pucci E}{1992}{\jpa}{25}{2631--2640}
\refn{refFE}
\refjl{Fujioka J and Espinosa A}{1980}{\JPSJ}{60}{4071--4075}

\refn{refCS}
\refjl{Cosgrove C M and Scoufis G}{1993}{\sam}{88}{25--87}
\refn{refFA}
\refjl{Fokas A S and Ablowitz M J}{1983}{\jmp}{23}{2033--2042}
\refn{refGromak}
\refjl{Gromak V I}{1975}{Diff. Eqns.}{11}{285--287}

\refn{refBurea}
\refjl{Bureau F}{1972}{Ann. Mat. Pura Appl. (IV)}{91}{163--281}
\refn{refBureb}
\refjl{Bureau F, Garcet A and Goffar J}{1972}{Ann. Mat. Pura Appl.
(IV)}{92}{177--191}
\refn{refChazy}
\refjl{Chazy J}{1911}{Acta Math.}{34}{317--385}
\refn{refCosa}
\refjl{Cosgrove C M}{1977}{\jpa}{10}{2093--2105}
\refn{refCosb}
\refjl{Cosgrove C M}{1978}{\jpa}{11}{2405--2430}
\refn{refJimbo}
\refjl{Jimbo M}{1982}{Publ. RIMS, Kyoto Univ.}{18}{1137--1161}
\refn{refJMii}
\refjl{Jimbo M and Miwa T}{1981}{Physica}{D2}{407--488}
\refn{refAir}
\refjl{Airault H}{1979}{Stud. Appl. Math.}{61}{31--53}
\refn{refGrob}
\refjl{Gromak V I}{1978}{Diff. Eqns.}{14}{1510--1513}
\refn{refLuka}
\refjl{Lukashevich N A}{1965}{Diff. Eqns.}{1}{561--564}
\refn{refMFokb}
\refjl{Mugan U and Fokas A S}{1992}{\jmp}{33}{2031--2045}
\refn{refOka}
\refjl{Okamoto K}{1987}{Funkcial. Ekvac.}{30}{305--332}
\refn{refLukb}
\refjl{Lukashevich N A}{1967}{Diff. Eqns.}{3}{994--999}
\refn{refMilne}
\refcf{Milne A E and Clarkson P A}{1993}{Applications of Analytic and
Geometric Methods to Nonlinear
Differential Equations}{Editor P.A.\ Clarkson}{{\it NATO ASI Series C:
Mathematical and Physical
Sciences\/}, Kluwer, Dordrecht}{pp341--352}
\refn{refKLM}
\refjl{Kitaev A V, Law C K and McLeod J B}{1993}{J. Diff. Int.
Eqns.}{}{to appear}
\refn{refOkb}
\refjl{Okamoto K}{1987}{Jap. J. Math.}{13}{47--76}
\refn{refGroc}
\refjl{Gromak V I}{1976}{Diff. Eqns.}{12}{740--742}
\refn{refLukc}
\refjl{Lukashevich N A}{1968}{Diff. Eqns.}{4}{1413--1420}

\def\cite#1{[#1]}
\def\DELTA{\Delta}%{\Delta\!^{(4)}}
\def\maple{{\sc maple}}
\def\dgb{{\sc dgb}}
\def\ft{{\d f\over \d t}}
\def\dgbs{{\sc dgbs}}
\def\sqr#1{\left(#1\right)^{1/2}}

\title{On a Shallow Water Wave Equation}
\author{Peter A. Clarkson and Elizabeth L. Mansfield}
\address{Department of Mathematics, University of Exeter, Exeter, EX4
4QE,
U.K.}
\bigskip
\jnl{Nonlinearity}
\date

\beginabstract
In this paper we study a shallow water equation derivable using the
Boussinesq approximation,
which includes as two special cases, one equation discussed by Ablowitz
\etal [{\it Stud.\ Appl.\
Math.}, {\bf53} (1974) 249--315] and one by Hirota and Satsuma [{\it
J.\ Phys.\ Soc.\ Japan}, {\bf40}
(1976) 611--612]. A catalogue of classical and nonclassical symmetry
reductions,
and a \p\ analysis, are given.  Of particular interest are families of
solutions found containing a
rich variety of qualitative behaviours. Indeed we exhibit and plot a
wide variety of solutions all of
which look like a two-soliton for $t>0$ but differ radically for $t<0$.
These families arise  as
nonclassical symmetry reduction solutions and solutions found using the
singular manifold method.
This example shows that nonclassical symmetries and the singular
manifold
method do not, in general, yield the same solution set. We also obtain
symmetry reductions of the
shallow water equation solvable in terms of solutions of the first,
third and fifth \p\ equations.

We give evidence that the variety of solutions found which exhibit
``nonlinear superposition'' is
not an artefact of the equation being linearisable since the equation
is solvable by inverse
scattering. These solutions have important implications with regard to
the numerical analysis for
the shallow water equation we study, which would not be able to
distinguish the solutions in an
initial value problem since an exponentially small change in the
initial conditions can result in
completely different qualitative behaviours.
 \endabstract
\vfill\eject
\pageno=1
\headline={\ifodd\pageno\rightheadline\else\leftheadline\fi}
\def\rightheadline{\tenrm\hfil {\sit Nonclassical symmetries and exact
solutions of a shallow water
wave equation}\hfil\folio}
\def\leftheadline{\tenrm\hfil {\sit Peter A Clarkson and Elizabeth L
Mansfield}\hfil\folio}

\section{Introduction}
In this paper we discuss the generalised shallow water wave ({\sc
gsww}) equation
$$\DELTA \equiv u_{xxxt} + \alpha u_x u_{xt} + \beta u_t u_{xx}
- u_{xt} - u_{xx} = 0,\eqn{eqgsww}$$
where $\alpha$ and $\beta$ are arbitrary, nonzero, constants.  This
equation, together with several
variants, can be derived from the classical shallow water theory in the
so-called Boussinesq
approximation \cite{\refWhitham}.
There are two special cases of this equation which have been discussed
in the
literature; (i), if $\alpha=\beta$
$$u_{xxxt} + \beta u_x u_{xt} + \beta u_t u_{xx}  - u_{xt} - u_{xx} =
0,\eqn{eqswwi}$$
which we shall call the {\sc swwi} equation,
and (ii) if $\alpha=2\beta$
$$u_{xxxt} + 2\beta u_x u_{xt} + \beta u_t u_{xx}  - u_{xt} - u_{xx} =
0,\eqn{eqswwii}$$
which we shall call the {\sc swwii} equation. These equations are often
written in the
nonlocal form (set $u_x=v$)
$$v_{xxt} + \beta v v_{t} - \beta v_{x}\partial_x^{-1}v_t  - v_{t} -
v_{x} =
0,\eqno(\eqswwi^*)$$ where $\left(\partial_x^{-1} f\right)(x) =
\int_x^\infty f(y)\,\d y$,
which was discussed by Hirota and Satsuma \cite{\refHS}, and
\def\eqswwa{$\eqswwi^*$}
\def\eqswwb{$\eqswwii^*$}
$$v_{xxt} + 2\beta v v_{t} - \beta v_{x}\partial_x^{-1}v_t  - v_{t} -
v_{x} =
0.\eqno(\eqswwii^*)$$
which was discussed by Ablowitz \etal \cite{\refAKNS} who
showed that it is solvable by inverse scattering (see \S4.2).
Furthermore Ablowitz \etal
\cite{\refAKNS} remark that (\eqswwb) reduces to the celebrated
Korteweg-de Vries ({\sc kdv})
equation
$$u_{t} + u_{xxx} + 6uu_x=0,\eqn{eqkdv}$$ which also is solvable by
inverse scattering
\cite{\refGGKM}, in the long wave, small amplitude limit. Equation
(\eqswwb) also has the desirable
properties of the regularized long wave ({\sc rlw}) equation
\cite{\refPer,\refBBM}
$$v_{xxt} +  v v_{x} - v_{t} - v_{x} = 0,\eqn{eqrlw}$$ sometimes called
the Benjamin-Bona-Mahoney
equation, in that it responds feebly to short waves. However, in
contrast to (\eqswwb), the {\sc
rlw} equation (\eqrlw) is thought not to be solvable by inverse
scattering (cf., \cite{\refMcLO}).

The {\sc gsww} equation (\eqgsww) is discussed by Hietarinta
\cite{\refHiet} who shows
that it can be expressed in Hirota's bilinear form \cite{\refHirt} if
and only if either (i),
$\alpha=\beta$, when it reduces to (\eqswwi), or (ii), $\alpha=2\beta$,
when it reduces to
(\eqswwii). Furthermore, as shown below, the {\sc gsww} equation
(\eqgsww) satisfies the necessary
conditions of the \p\ tests due to Ablowitz \etal
\cite{\refARSa,\refARSb} (see \S2) and Weiss \etal
\cite{\refWTC} (see \S4.1) to be completely integrable if and only if
$\alpha=\beta$ or
$\alpha=2\beta$. We show in \S4.2 that the {\sc gsww} equation
(\eqgsww) is solvable by
inverse scattering techniques in these two special cases. These results
strongly suggest that the
{\sc gsww} equation is completely integrable if and only if it has one
of the two special forms
({\eqswwi}) or ({\eqswwii}).

The {\sc swwi} equation ({\eqswwi}) and {\sc swwii} equation
({\eqswwii}) arise as a reduction of
several higher-dimensional \pdes\ which have been discussed in the
literature.
The {\sc swwi} equation ({\eqswwi}) arises as a reduction of:
\item{1. } The $2+1$-dimensional equation
$$u_{yt} + u_{xxxy} - 3u_{xx}u_y- 3u_xu_{xy}=0,\eqn{eqblmp}$$ which
reduces to the {\sc kdv}
equation (\eqkdv) if $y=x$. Boiti \etal \cite{\refBLMP} developed an
inverse scattering scheme to
solve the Cauchy problem for (\eqblmp), for initial data decaying
sufficiently rapidly at infinity;
this was formulated as a nonlocal Riemann-Hilbert problem.
\item{2. } The $3+1$-dimensional equation
$$u_{yt} + u_{xxxy} -3u_{xx}u_y- 3u_xu_{xy} -u_{xz}=0,\eqn{eqjm}$$
which was introduced by Jimbo and
Miwa \cite{\refJM} as the second equation in the so-called
Kadomtsev-Petviashvili hierarchy of
equations; however (\eqjm) is not completely integrable in the usual
sense (see \cite{\refDGRW}).
\item{3. } The $2+1$-dimensional
equation
$$u_{tt} - u_{xx} - u_{yy} + u_xu_{xt}+u_yu_{yt} - u_{xxtt} -
u_{yytt}=0,\eqn{eqkt}$$ which was
introduced by  Yajima \etal \cite{\refYOS} as a model of ion-acoustic
waves in plasmas; Kako and
Yajima \cite{\refKY} have studied ``soliton interactions'' for
(\eqkt).

The {\sc swwii} equation ({\eqswwii}) arises as a reduction of the
$2+1$-dimensional
equation
$$u_{xt} + u_{xxxy} - 2u_{xx}u_y - 4u_xu_{xy}=0,\eqn{eqbogii}$$ which,
like ({\eqblmp}), reduces to
the {\sc kdv} equation (\eqkdv) if $y=x$, though note that the term
$u_{yt}$ in ({\eqblmp}) is
replaced by $u_{xt}$ in (\eqbogii). Bogoyaviemskii
\cite{\refBogi,\refBogii} discusses the inverse
scattering method of solution for (\eqbogii).

In \S\S2 and 3, we find first the classical Lie group of symmetries and
associated reductions of
(\eqgsww) and then nonclassical symmetries and reductions of
(\eqgsww).
The classical method for finding symmetry reductions of \pdes\ is the
Lie group method of
infin\-ites\-imal transformations (cf., \cite{\refBK,\refOlver}).
Though this
method is entirely algorithmic, it often involves a large amount of
tedious algebra and auxiliary
calculations  which can become virtually unmanageable if attempted
manually, and so symbolic
manipulation programs have  been developed, for example in {\sc
macsyma}, {\sc maple},  {\sc
mathematica}, {\sc mumath} and {\sc reduce}, to facilitate the
calculations. A survey of the
different packages presently  available and a discussion of their
strengths and applications is
given by Hereman \cite{\refHere}. In this paper we use the {\sc
macsyma} program {\tt
symmgrp.max} \cite{\refCHW}.

In recent years the nonclassical method due Bluman and Cole
\cite{\refBCa} (in the sequel
referred to as the {\it nonclassical method\/}, see \S3 for further
details), which is also known as
the ``method of conditional symmetries'' \cite{\refLW} or the ``method
of partial symmetries of the
first type'' \cite{\refVor},  and the direct method of Clarkson and
Kruskal \cite{\refCK} have been
used to generate many new symmetry reductions and exact solutions for
several physically significant
\pdes\ that are {\it not\/} obtainable using the classical Lie method,
which represents important
progress (see for example \cite{\refPACrev,\refFush} and references
therein).  Since solutions of
\pdes\ asymptotically tend to solutions of lower-dimensional equations
obtained by symmetry
reduction, some of these special solutions will illustrate important
physical phenomena. In
particular, exact solutions arising from symmetry methods can often be
effectively used to study
properties such as asymptotics and ``blow-up''
(cf.\ \cite{\refGalaka,\refGDEKS}). Furthermore,
explicit solutions (such as those found by symmetry methods) can play
an important role in the
design and testing of numerical integrators; these solutions provide an
important practical check on
the accuracy and reliability of such integrators
(cf.\ \cite{\refAmesii,\refShok}).

There is much current interest in the determination of symmetry
reductions of \pdes\ which reduce
the equations to \odes. Often one then checks if the resulting \ode\ is
of {\it\p\ type\/}, i.e.,
its solutions have no movable singularities other than poles. It
appears to be the case that
whenever the \ode\ is of \p\ type then it can be solved explicitly,
leading to exact solutions to
the original equation. Conversely, if the resulting \ode\ is not of
\p\ type, then often one is
unable to solve it explicitly.

The method used to find solutions of the determining equations for the
infinitesimals in both the
classical and nonclassical case is that of Differential Gr\"obner Bases
(\dgbs), defined to be a
basis ${\cal B}$ of the differential ideal generated by the system such
that every member of the
ideal pseudo-reduces to zero with respect to
${\cal B}$.  This method provides a systematic framework for finding
integrability and compatibility
conditions of an overdetermined system of partial differential
equations. It avoids the problems of
infinite loops in reduction processes, and yields, as far as is
currently possible, a
``Triangulation" of the system from which the solution set can be
derived more easily \cite{\refReida--\refCMa}. In a sense, a
\dgb\ provides the maximum amount
of information possible using elementary differential and algebraic
processes in a finite time.

In pseudo-reduction, one is allowed to multiply the expression being
reduced by differential, that
is, non-constant, coefficients of the highest derivative terms of the
reducing equations. The reason
one must do this is that on nonlinear systems, the algorithms for
calculating the differential
analogue of a Gr\"obner Basis will not terminate if only strict
reduction is allowed.  What this
means in practice is that such coefficients are assumed to be nonzero.
To obtain solutions of the system that
evaluate to zero one of these coefficients, one needs to include it
with the system from the start
of the calculation.  Such a solution is called a singular integral for
the obvious reason
(cf., \cite{\refZwil}).

The major problems with the \dgb\ method in practice are its poor
complexity and expression swell.
However, on systems where the process can be completed within
reasonable limits, by which is meant
that the length of the expressions obtained is small enough to be
meaningful,
 the output is extremely useful.  Comparing the determining equations
 for classical symmetries and a
triangulation for that system illustrates this point; see (2.7) and
(2.8) below.  For
nonlinear systems, \dgbs\ have been used effectively to solve the
determining equations for nonclassical symmetries
\cite{\refCMa,\refCMb}, using various strategies
which address the complexity problem and which minimize the number of
singular integral cases to be
considered, i.e. which minimise the differential coefficients used in
the pseudo-reduction processes.

A much older method of finding a basis for the ideal of a system from
which
formal solutions may be derived, due to Janet, has been implemented for
linear systems
\cite{\refSchw,\refTop}. Also for linear systems (and linear
differential-difference systems),
 the differential analogue of Buchberger's algorithm \cite{\refBuchi}
 for calculating an
algebraic Gr\"obner Basis has been implemented \cite{\refPank}. For
orthonomic systems, those whose
members are solvable for their leading derivative term, the
Reid-Wittkopf Differential Algebra
package \cite{\refRW} will calculate the Standard Form of the system,
the
number of arbitrary constants and functions a formal solution depends
on (see also
\cite{\refSchw}), and the formal power series solution to any order
\cite{\refReidb}.  This package handles equations with nontrivial
coefficients of the leading
derivative terms provided {\sc maple} can solve the expression
(algebraically) for the leading
term.  One can then systematically go through the singular integrals
using the {\tt divpivs} command.

The triangulations of the systems of determining equations for
infinitesimals arising in the
classical and nonclassical methods in this article were all performed
using the
\maple\ package {\tt diffgrob2} \cite{\refMD}. This package was written
specifically to handle fully
nonlinear equations.  All calculations are strictly ``polynomial", that
is, there is no division.
Implemented there are the Kolchin-Ritt algorithm, the differential
analogue of Buchberger's
algorithm using pseudo-reduction instead of reduction, and extra
algorithms needed to calculate a
\dgb\  (as far as possible using the current theory), for those cases
where the Kolchin-Ritt
algorithm is not sufficient \cite{\refMF}.  Designed to be used
interactively as well as
algorithmically, the package has proved useful for solving some fully
nonlinear systems.  As yet,
however, algorithmic methods for finding the most efficient orderings,
the best method of choosing
the sequence of pairs to be cross-differentiated, for deciding when to
integrate and read off
coefficients of independent functions in one of the variables, for
finding the best change of
coordinates, and so on, are still the subject of much investigation.

The nonclassical symmetry reductions obtained for (\eqswwi) generate a
wide variety of interesting
exact analytical solutions of the equations which we plot (using {\sc
maple}) in the
Figures.  In \S4 we apply the \p\ test due to Weiss \etal
\cite{\refWTC} to (\eqgsww), and then
obtain another family of solutions of (\eqswwi) using the singular
manifold method
\cite{\refWTC,\refWeiss}.  We then discuss the  scattering problems for
(\eqswwi) and (\eqswwii) and
show how the arbitrary functions in the solutions obtained would appear
in a solution of (\eqswwi)
obtained by the inverse scattering method.  Finally in \S5 we discuss
our results.

\section{Classical Symmetries}
To apply the classical method to the {\sc gsww} equation ({\eqgsww}) we
consider the
one-parameter Lie group of infinitesimal transformations in
$(x,t,u)$ given by
$$\eqalignno{
\~{x} &={x}+ \varepsilon {\xi}(x,t,u) + O(\varepsilon^2),
&\eqnm{eqinftr}{a}\cr
\~{t} &= {t} + \varepsilon {\tau}(x,t,u) + O(\varepsilon^2), &\eqnr{b}
\cr
\~{u} &= {u} + \varepsilon {\phi}(x,t,u) + O(\varepsilon^2),
&\eqnr{c}\cr}$$
where $\varepsilon$ is the group parameter. Then one requires that
this transformation leaves invariant the set
$${\cal S}_{\DELTA} \equiv \left\{u(x,t):
\DELTA=0\right\},\eqn{Sdelta}$$ of
solutions of ({\eqgsww}). This yields an overdetermined, linear system
of equations for the
infin\-ites\-imals ${\xi}(x,t,u)$, ${\tau}(x,t,u)$ and ${\phi}(x,t,u)$.
The associated Lie algebra
of infin\-ites\-imal symmetries is the set of vector fields of the form
$$ {\bf v} = \xi(x,t,u){\partial\over\partial x} +
\tau(x,t,u){\partial\over\partial t} +
\phi(x,t,u){\partial\over\partial u}. \eqn{eqvf}$$
Having determined the infinitesimals, the symmetry variables are found
by solving the
characteristic equations
$${{\d}x \over\xi(x,t,u)} = {{\d}t \over\tau(x,t,u)} = {{\d}u
\over\phi(x,t,u)},\eqn{chareq}$$
which is equivalent to solving the invariant surface condition
$$\psi\equiv\xi(x,t,u)u_x +\tau(x,t,u)u_t - \phi(x,t,u)=0.\eqn{insc}$$

The set ${\cal S}_{\DELTA}$ is invariant under the transformation
(\eqinftr) provided that
$$\left.\pr4{\bf
v}\left(\DELTA\right)\right|_{\DELTA=0}=0,\eqn{privth}$$
where $\pr4{\bf v}$ is the fourth prolongation of the vector field
(\eqvf), which is
given explicitly in terms of ${\xi}$, ${\tau}$ and ${\phi}$ (cf.,
\cite{\refOlver}).
This yields the following fourteen determining equations,
$$\eqalignno{
&\tau_{u}=0,\qquad\tau_{x}=0,\qquad\xi_{u}=0,\qquad \xi_{t}=0,
&\eqnm{cldeteqs}{a}\cr
&\phi_{uu}=0,\qquad\phi_{tu}=0,\qquad\phi_{xu}-\xi_{xx}=0,\qquad\phi_{u}+\xi_{x}=0,
&\eqnr{b}\cr
&\beta \phi_{t}-\tau_{t}-\xi_{x}=0,\qquad
2\beta\phi_{xu}+\alpha\phi_{xu}-\beta
\xi_{xx}=0, &\eqnr{c}\cr
&\beta\phi_{xx}+\phi_{xxxu}-\phi_{xu}=0,\qquad \alpha \phi_{xt} - 2
\phi_{xu} + \xi_{xx}=0,
&\eqnr{d}\cr &\phi_{xxxt} - \phi_{xx} - \phi_{xt}=0,\qquad
\alpha \phi_{x} + 3 \phi_{xxu} - \xi_{xxx} - 2 \xi_{x}=0.
&\eqnr{e}\cr}$$ These equations were
calculated using the {\sc macsyma} package {\tt symmgrp.max}
\cite{\refCHW}.

A triangulation or standard form \cite{\refReida--\refCMa} (see also
\cite{\refSchw,\refTop}) of the
determining equations ({\cldeteqs}) for classical symmetries of the
{\sc gsww} equation ({\eqgsww})
is  the following system of eight equations,
$$\eqalignno{&\xi_{u}=0,\qquad \xi_{t}=0,\qquad \xi_{xx}=0,
\qquad\tau_{u}=0,\qquad \tau_{x}=0,&\eqnm{}{a}\cr
&\alpha\phi_{x}-2\xi_{x}=0,\qquad
\beta\phi_{t}-\tau_{t}-\phi_{x}=0,\qquad
\phi_{u}+\xi_{x}=0,&\eqnr{b}\cr}$$
from which we easily obtain the following infinitesimals,
$$ \xi=\cc1 x+\cc2,\qquad \tau=g(t),\qquad
\phi=-\cc1
\left[u-{2x\over\alpha}-{t\over\beta}\right]+{g(t)\over\beta}+\cc3,
\eqn{clinfs}$$
where $g(t)$ is an arbitrary function and $\cc1$, $\cc2$ and $\cc3$ are
arbitrary constants.
The associated vector fields are:
$${\bf v}_1 = x{\partial\over\partial x} -
\left(u-{2x\over\alpha}-{t\over\beta}\right){\partial\over\partial
u},\qquad {\bf v}_2 =
{\partial\over\partial x},\qquad {\bf v}_3 = {\partial\over\partial
u},\qquad {\bf v}_4(g) =
g(t)\left({\partial\over\partial t} +
{1\over\beta}{\partial\over\partial u}\right).$$
We remark that ${\bf v}_4(g)$ shows that ({\eqgsww}) is invariant under
the following variable
coefficient ``Galilean transformation''
$$\~x=x,\qquad \~t= g(t),\qquad \~u=u+[g(t)-t]/\beta,\eqn{eqGalTr}$$
i.e., if $u(x,t)$ is a
solution of ({\eqgsww}), then so is $\~u(\~x,\~t)$.
Solving ({\chareq}), or equivalently solving ({\insc}), we obtain two
canonical symmetry reductions.

\smallskip\noindent{\bf Case 2.1 $\cc1\ne0$}. In this case we set
$\displaystyle {1\over g(t)} =
{1\over f(t)}{{\d}f\over{\d t}}$,
$\cc1=1$ and $\cc2=\cc3=0$  and obtain the symmetry reduction
$$ u(x,t) = f(t) w(z) + {x\over \alpha} + {t\over\beta},\qquad z=xf(t),
\eqn{clsri} $$
where $w(z)$ satisfies
$$ z\wzzzz+4\wzzz+ (\alpha+\beta)z\wz\wzz + \beta w\wzz +
2\alpha\left(\wz\right)^2=0. \eqn{clsriic}
$$
It is straightforward to show using the algorithm of Ablowitz \etal
\cite{\refARSb} that this equation is of \p-type only if either (i),
$\alpha=\beta$ or (ii),
$\alpha=2\beta$; in the Appendix it is shown that in these two special
cases (\clsriic) is solvable in terms of
solutions of the third \p\ equation \cite{\refInce}
$${\d^2y\over\d x^2} = {1\over y}\left({\d y\over\d x}\right)^2-
{1\over x}{\d y\over\d
x} + a y^3 + {b y^2+c\over x} + {d\over y},\eqn{eqpiii}$$
and the fifth \p\ equation,
$${\d^2y\over\d x^2} = \left\{{1\over 2y}+{1\over y-1}\right\}\left({\d
y\over\d x}\right)^2-
{1\over x}{\d y\over\d x}+ {(y-1)^2\over x^2}\left\{a y + {b\over
y}\right\} + {cy\over x} +
{dy(y+1)\over y-1},\eqn{eqpv}$$ with $a$, $b$, $c$ and $d$ constants.
Hence the \p\ Conjecture \cite{\refARSa,\refARSb} predicts that a
necessary condition for (\eqgsww)
to be completely integrable is that $(\alpha-\beta)(\alpha-2\beta)=0$,
i.e., only if (\eqgsww) has
one of the two special forms (\eqswwi) or (\eqswwii). We remark that
the occurrence of the third and
fifth \p\ equations is slightly surprising since for the Boussinesq
equation
$$ u_{xx} + 3(u^2)_{xx} + u_{xxxx}=u_{tt},\eqn{}$$
symmetry reductions reduce the equation to the first, second and fourth
\p\ equations
\cite{\refCK}.

\smallskip\noindent{\bf Case 2.2 $\cc1=0$}. In this case we set
$\displaystyle {1\over g(t)} =
{{\d}f\over{\d t}}$, $\cc2=1$ and $\cc3=-1/\beta$ and obtain the
symmetry reduction
$$  u(x,t) = w(z) +  {t/\beta},\qquad z=x-f(t), \eqn{clsrii}
$$ where $w(z)$ satisfies
$$ \wzzzz + (\alpha+\beta)\wz\wzz=0. \eqn{clsriib}$$
Setting $W=\d w/\d z$ and integrating twice yields
$$ \left(\d W\over\d z\right)^2 + \tfr13(\alpha+\beta)W^3= AW+B,
\eqn{clsriid}$$
where $A$ and $B$ are constants of integration. This equation is
equivalent to the Weierstrass
elliptic function equation
$$ \left(\d\wp\over\d z\right)^2 = 4\wp^3-g_2\wp-g_3, \eqn{clsriie}
$$
where $g_2$ and $g_3$ are arbitrary constants \cite{\refWW}.

\section{Nonclassical Symmetries}
There have been several generalisations of the classical Lie group
method for symmetry reductions.
Bluman and Cole \cite{\refBCa}, in their study of symmetry reductions
of
 the  linear heat equation, proposed the so-called nonclassical method
 of group-invariant solutions.
This method involves considerably more algebra and associated
calculations than the classical Lie
method. In fact, it has been  suggested that for some
\pdes, the calculation of these nonclassical reductions might be too
difficult  to do explicitly,
especially if attempted manually since the associated determining
equations are now an
overdetermined, {\it nonlinear\/} system. For some equations such as
the {\sc kdv} equation (\eqkdv), the nonclassical method does not yield
any additional symmetry
reductions to those obtained using the classical Lie method, while
there are
\pdes\ which do possess symmetry reductions {\it not\/} obtainable
using the classical Lie
group method. It should be emphasised that the associated vector fields
arising from the
nonclassical method do not form a vector space, still less a Lie
algebra, since the invariant
surface condition (\insc) depends upon the particular reduction.

In the nonclassical method one requires only the subset of ${\cal
S}_{\DELTA}$ given by
$${\cal S}_{\DELTA,\psi} = \left\{u(x,t):\DELTA(u)
=0,\psi(u)=0\right\},\eqn{nclset}$$
where ${\cal S}_{\DELTA}$ is as defined in (\Sdelta) and $\psi=0$ is
the invariant surface
condition (\insc), is invariant under the transformation ({\eqinftr}).
The usual method of applying
the nonclassical method (e.g., as described in \cite{\refLW}), to the
{\sc gsww} equation (\eqgsww)
involves applying the prolongation $\pr4{\bf v}$ to the system of
equations given by (\eqgsww) and
the invariant surface condition (\insc) and requiring that the
resulting expressions vanish for
$u\in{\cal S}_{\DELTA,\psi}$, i.e.,
$$\left.\pr4{\bf
v}\left(\DELTA\right)\right|_{\DELTA=0,\psi=0}=0,\qquad
\left.\pr1{\bf v}(\psi)\right|_{\DELTA=0,\psi=0}=0.\eqn{plpde}$$  It is
easily shown that
$$\pr1{\bf v}(\psi)=-\left(\xi_uu_x+\tau_uu_t-\phi_u\right)\psi,$$
which vanishes identically when
$\psi=0$ without imposing any conditions upon
$\xi$, $\tau$ and $\phi$. However as shown in \cite{\refCMb}, this
procedure for applying the
nonclassical method can create difficulties, particularly when
implemented in symbolic
manipulation programs. These difficulties often arise for equations
such as (\eqgsww)  which require
the use of differential consequences of the invariant surface condition
(\insc). In \cite{\refCMb}
we proposed an algorithm for calculating the determining equations
associated with the nonclassical
method which avoids many of the difficulties commonly encountered; we
use that algorithm here.

In the canonical case when $\tau\not\equiv0$ we set $\tau=1$. We omit
the special
case $\tau\equiv0$; in that case one obtains a single condition for
$\phi$ with 424 summands, and even considering the subcase $\phi_u=0$
leads to an equation more complex than the one we are studying.
Eliminating $u_t$, $u_{xt}$ and $u_{xxxt}$, in (\eqgsww) using the
invariant surface condition
(\insc) yields
$$\eqalignno{{\bf\tilde{\hbox{$\Delta$}}}\equiv
\phi_{u}&u_{xxx}+3\phi_{uu}u_x
u_{xx}+3\phi_{ux}u_{xx} +\phi_{uuu}u_x^3+3\phi_{xuu}
u_x^2+3\phi_{xxu}u_x +\phi_{xxx} \cr
&-\xi_{xxx}u_x
-3\xi_{xx}u_{xx}-3\xi_{x}u_{xxx}-\xi_{uuu}u_x^4-3\xi_{xuu}u_x^3-6\xi_{uu}u_{x}^2u_{xx}\cr
&-3\xi_{xxu}u_x^2-9\xi_{xu}u_x
u_{xx}-\xi_{u}\left(4u_xu_{xxx}+3u_{xx}^2\right)-\xi u_{xxxx}\cr
&+\left(\alpha u_x-1\right)\left[\phi_x + \phi_uu_x-\xi_x u_x-\xi_u
u_x^2-\xi u_{xx}\right]
+u_{xx}\left[\beta \left(\phi-\xi u_x\right)-1\right]
=0,&\eqnn{swwisc}\cr}$$
with $t$ a parameter, which involves the infinitesimals $\xi$ and
$\phi$ that are to be determined.
Now we apply the classical Lie algorithm to this equation using the
fourth prolongation $\pr4{\bf
v}$ and eliminate $u_{xxxx}$ using ({\swwisc}). This yields the
following overdetermined, nonlinear
system of equations for ${\xi}$ and
${\phi}$ (contrast the classical case discussed in \S2 above where they
are linear).
$$\eqalignno{
				    &\xi_{u}=0,&\eqnm{noncldeqs}{i}\cr
				   &\phi_{uu}=0,&\eqnr{ii}\cr
			 &(\alpha + \beta) (\phi_{u} +
			 \xi_{x})=0,&\eqnr{iii}\cr
      &\xi \phi_{tu}  + 3 \xi^{2} \xi_{xx} + 3 \xi_{t} \xi_{x} - 3 \xi
      \xi_{xt} - 3 \xi^{2}
\phi_{xu} - \xi_{t} \phi_{u}=0,&\eqnr{iv}\cr
     & \alpha \xi \phi_{u}^{2} - \alpha \xi_{t} \phi_{u} + \alpha \xi
     \phi_{tu} - 2 \beta \xi^{2}
\phi_{xu} - \alpha \xi^{2} \phi_{xu} + \beta \xi^{2} \xi_{xx}
  - \alpha \xi \xi_{x}^{2} + \alpha \xi_{t} \xi_{x} - \alpha \xi
  \xi_{xt}=0,&\eqnr{v}\cr
  & 3 \xi \phi_{xu} \phi_{xx} + \beta \xi \phi \phi_{xx} - \xi_{t}
  \phi_{xxx}
       - 3 \xi \xi_{xx} \phi_{xx} - \xi \phi_{xx} + \alpha \xi
       \phi_{x}^{2}
       + 3 \xi \phi_{xxu} \phi_{x} - \xi \xi_{xxx} \phi_{x}\cr
       &\qquad- 2 \xi \xi_{x} \phi_{x} + \xi_{t} \phi_{x} + \xi \phi
       \phi_{xxxu}
 - \xi \phi \phi_{xu} + \xi \phi_{xxxt} - \xi \phi_{xt}=0,&\eqnr{vi}\cr
 & 3 \xi \phi_{u} \phi_{xu} - 6 \xi \xi_{x} \phi_{xu} - 3 \xi_{t}
 \phi_{xu} - \alpha \xi^{2}
\phi_{x} - 3 \xi^{2} \phi_{xxu} + \beta \xi \phi \phi_{u}
  - 3 \xi \xi_{xx} \phi_{u} + 3 \xi \phi_{xtu} + \beta \xi \phi_{t}\cr
      &\qquad+ \beta \xi \xi_{x} \phi - \beta \xi_{t} \phi + \xi^{2}
      \xi_{xxx}
      + 6 \xi \xi_{x} \xi_{xx} + 3 \xi_{t} \xi_{xx} + 2 \xi^{2} \xi_{x}
      - \xi \xi_{x} - 3 \xi \xi_{xxt} + \xi_{t}=0,&\eqnr{vii}\cr &9 \xi
      \xi_{xx} \phi_{xu} + \beta
\xi^{2}
\phi_{xx} + \alpha \xi_{t} \phi_{x}
     + 3 \xi \xi_{x} \phi_{xxu} - 2 \alpha \xi \phi_{u} \phi_{x}
 - 6 \xi \phi_{xu}^{2} - \alpha \xi \phi_{xt} - \xi \xi_{x}
 \xi_{xxx}\cr
      &\qquad+ 2 \xi \phi_{xu} + \xi \xi_{xxx} \phi_{u} - 3 \xi
      \phi_{u} \phi_{xxu}
     - 3 \xi \xi_{xx}^{2} - 2 \beta \xi \phi \phi_{xu} - 3 \xi
     \phi_{xxtu}
 + 2 \xi \xi_{x} \phi_{u} - 2 \xi \xi_{x}^{2} - \xi^{2} \phi_{xu}\cr
      &\qquad- \xi_{t} \phi_{u} + \xi \phi_{tu} + \xi_{t} \xi_{x} - \xi
      \xi_{xt}
 + \beta \xi \xi_{xx} \phi + \xi^{2} \phi_{xxxu} - \alpha \xi \phi
 \phi_{xu}\cr
      &\qquad- \xi_{t} \xi_{xxx} - \xi \xi_{xx} + \xi \xi_{xxxt}
     + 3 \xi_{t} \phi_{xxu}=0.&\eqnr{viii} \cr}$$
These equations were calculated using the {\sc macsyma} package {\tt
symmgrp.max} \cite{\refCHW}.
We then used the method of {\sc dgbs} as outlined in
\cite{\refCMa,\refCMb} to solve this system.

\smallskip\noindent{\bf Case 3.1 $\alpha+\beta\ne 0$}.
In this case it is straightforward to obtain the condition
$$ \xi_{xx} \xi^{2} (\alpha+\beta) (3 \beta - 2 \alpha)=0.$$
The case $3 \beta - 2 \alpha=0$ leads to no solutions different from
those
obtained using the classical method.

\smallskip\noindent{\sl Subcase 3.1.1  $\xi_x\ne 0$}.
This is the generic case which has the solution
$$
\xi = {[\cc1 x + \cc2]f(t)},\qquad
\phi =- {\cc1f(t)}\left[u - {2 x\over\alpha} + {\cc2 -
t\over\beta}\right] + {1\over\beta},
$$
where $f(t)$ is an arbitrary function and $\cc1\not=0$ and $\cc2$ are
arbitrary constants. These
are equivalent to the infinitesimals ({\clinfs}) obtained using the
classical method.

\smallskip\noindent{\sl Subcase 3.1.2 $\xi_x= 0$}.
In this case it is easy to obtain the condition
$$\phi_{xx}\xi^3(\beta-\alpha)(\alpha+\beta)=0.$$
There are two subcases to consider.

\noindent(i) $\alpha\ne\beta$, $\xi_x=0$. In this case the solution is
$$\xi= {f(t)},\qquad \phi=  {\cc{3}f(t)}+{1/\beta},$$
where $f(t)$ is an arbitrary function and $\cc3$ is an arbitrary
constant, which is equivalent to
the infinitesimals ({\clinfs}) obtained using the classical method in
the case when $\cc1=0$.

\noindent(ii) $\alpha= \beta$, $\xi_x=0$.
In this case, we obtain the following \dgb\ for $\xi$, $\phi$
$$\eqalignno{
&\xi_u=0,\qquad
\xi_x=0,\qquad
\phi_u=0,&\eqnn{EQiii}\cr
&\xi\phi_{xxxx}-(\xi+1)\phi_{xx}+\beta\phi\phi_{xx}+\beta\phi_x^2=0,&\eqnn{EQiv}\cr
&\beta\xi^2\phi_x-\beta\xi\phi_t+\beta\xi_t\phi-\xi_t=0. &\eqnn{EQv}
\cr}$$
Thus $\xi$ is an arbitrary function of $t$, and so we set $\xi(t)=\d
f/\d t$.  It is easiest to
integrate ({\EQv}) first using the method of characteristics which
yields
$$ \phi=2V(\zeta)\ft +{1\over\beta},\qquad \zeta =x+f(t).\eqn{weqn}$$
Equation ({\EQiv}) can be integrated twice with respect to $x$.  This
yields
$$ \ft \phi_{xx}+\tfr12\beta\phi^2-\left[\ft
+1\right]\phi=x\lambda(t)+\mu(t),\eqn{aPi}$$
for some arbitrary functions $\lambda(t)$ and $\mu(t)$.
Substituting ({\weqn}) into ({\aPi}) yields
$$2\left(\ft\right)^2\left[{\d^2V\over\d\zeta^2}+\beta V^2-V\right]
=\lambda(t)[\zeta -f(t)]+\mu(t)+{1\over\beta}\ft+{1\over 2\beta}=0.$$
We obtain an \ode\ for $V(\zeta)$ if we set
$$\lambda(t)=2\cc{4}\left(\ft\right)^2,\qquad
\mu(t)=2\cc{4}f(t)\left(\ft\right)^2-{1\over\beta}\ft -{1\over 2\beta}
+\cc5,$$
where $\cc{4}$ and $\cc{5}$ are arbitrary constants, yielding
$${\d^2V\over\d\zeta^2}+\beta V^2-V = \cc{4}\zeta+\cc5.\eqn{eqpi}$$
This equation is equivalent to the first Painlev\'e equation
\cite{\refInce}
$${\d^2y\over\d x^2} = 6y^2 + x,\eqn{pieq}$$
if $\cc{4}\ne0$, otherwise it is equivalent to the Weierstrass elliptic
function
equation ({\clsriie}). Therefore we obtain the infinitesimals
$$\xi = \ft,\qquad \phi = 2V(\zeta) \ft +{1\over\beta},\eqn{}$$
where $\zeta=x+f(t)$, $f(t)$ is an arbitrary function and $V(\zeta)$
satisfies (\eqpi).

Hence solving the characteristic equations (\chareq) yields the
nonclassical symmetry reduction
$$ u(x,t) = v(\zeta)+w(z)+t/\beta,\qquad \zeta=x+f(t),\quad
z=x-f(t),\eqn{eqstar}
$$ where $f(t)$ is an arbitrary function and
$v(\zeta)=\int^\zeta_{-\infty} V(\zeta_1)\,\d\zeta_1$
and
$w(z)$ satisfy
$$ \vzzzz + 2\beta \vz\vzz - \vzz = - \lambda,\eqno\eqnm{vweq}{a}$$ and
$$ \wzzzz + 2\beta \wz\wzz - \wzz = \lambda,\eqno\eqnr{b}$$
respectively, with
$\lambda$ a ``separation'' constant. Integrating ({\vweq}) and setting
$V=\d v/\d\zeta$ and
$W=\d w/\d z$, yields
$$ {\d^2V\over\d \zeta^2}  + \beta V^2 -V = - \lambda \zeta +
\mu_1,\eqno\eqnm{VWeq}{a}$$ and
$$ {\d^2W\over\d z^2} + \beta W^2 - W = \lambda z+
\mu_2,\eqno\eqnr{b}$$  respectively, where
$\mu_1$ and $\mu_2$ are arbitrary constants. If $\lambda\not=0$ then
these equations are equivalent
to the first Painlev\'e equation ({\pieq}), whilst if $\lambda=0$ then
they are equivalent to
the Weierstrass elliptic function equation ({\clsriie}).

In particular, if $\lambda=0$, then equations ({\VWeq}) possess the
special solutions
$$ V(\zeta) = {6\kappa_1^2\over\beta}\sech^2\left(\kappa_1\zeta\right)
+{1-\sqr{1+4\mu_1\beta}\over2\beta},
\qquad W(z) = {6\kappa_2^2\over\beta}\sech^2\left(\kappa_2
z\right)+{1-\sqr{1+4\mu_2\beta}\over2\beta},
$$ where $\kappa_1=\tfr12\left(1+4\mu_1\beta\right)^{1/4}$ and
$\kappa_2=\tfr12\left(1+4\mu_2\beta\right)^{1/4}$. \ Hence we obtain
the exact solution of
({\eqswwi}) given by
$$\eqalignno{ u(x,t) =
{6\kappa_1\over\beta}&\tanh\left\{\kappa_1\left[x+f(t)\right]\right\} +
{6\kappa_2\over\beta}\tanh\left\{\kappa_2\left[x-f(t)\right]\right\}\cr&+
{x(1-2\kappa_1^2-2\kappa_2^2)\over\beta} +
{2f(t)(\kappa_2^2-\kappa_1^2)\over\beta} +
{t\over\beta},&\eqnn{swwsol}\cr}
$$ where $f(t)$ is an arbitrary function.

If $\mu_1=\mu_2=0$ then $\kappa_1=\kappa_2=\tfr12$ and (\swwsol)
simplifies to
$$ u(x,t) = {3\over\beta}\tanh\left\{\tfr12\left[x+f(t)\right]\right\}
+
{3\over\beta}\tanh\left\{\tfr12\left[x-f(t)\right]\right\}+
{t\over\beta}.\eqn{swwsoli}$$ In Figures
1 and 2 we plot $u_x$ with $u$ given by (\swwsoli) for various choices
of the arbitrary function
$f(t)$. This is one of the simplest, nontrivial family of
solutions of (1.1) with $\alpha=\beta$, using this reduction. In Figure
1,  $f(t)$ is chosen so that
$f(t)\sim t+t_0$, as $t\to\infty$, where
$t_0$ is a constant. Consequently all the solutions plotted in Figure 1
have a similar asymptotic
behaviour as $t\to\infty$. However the  asymptotic behaviours as
$t\to-\infty$ are radically
different.

In the special case when $f(t)=ct$, then choosing
$\kappa_1=\tfr12(1+1/c)^{1/2}$ and
$\kappa_2=\tfr12(1-1/c)^{1/2}$ in (\swwsol) yields the two-soliton
solution of ({\eqswwi}) given by
$$ u(x,t) = {3\over\beta}\left\{\sqr{c+1\over c}
\tanh\left[\sqr{c+1\over 4c}(x+ct)\right] + \sqr{c-1\over c}
\tanh\left[\sqr{c-1\over 4c}(x-ct)\right]\right\}.\eqn{swwsolii}$$
This solution is of special interest since such two-soliton solutions
are normally associated with
so-called Lie-B\"acklund transformations (cf., \cite{\refAndIb})
whereas (\swwsolii) has arisen from
a Lie point symmetry, albeit nonclassical. A plot of (\swwsolii) for
$c=2$ is given in Figure 3a
where it is compared to the so-called resonant two-soliton solution
obtained using the singular
manifold method in \S4.1 below.

We remark that this ``decoupling'' of the nonclassical symmetry
reduction solution (\eqstar) into a
function of $\zeta=x+f(t)$ and a function of $z=x-f(t)$ occurs for the
{\sc gsww} equation
(\eqgsww) only in this special case when $\alpha=\beta$.

\smallskip\noindent{\bf Case 3.2 $\alpha+\beta=0$}.
Substituting $\phi=u\theta(x,t)+\sigma(x,t)$ it is easy to find the
condition
$$\theta\theta_{xx}-\theta_x^2=0.$$
Thus either $\theta_x=0$ or
$\theta(x,t)=\exp\{x\lambda_1(t)+\lambda_2(t)\}$, where
$\lambda_1(t)$ and $\lambda_2(t)$ are arbitrary functions.  In fact,
unless
$\theta_x=0$ there are no solutions.  This can be shown by substituting
into the equations
the expressions $\theta_x=\theta\lambda_1$ and
$\theta_t=(x\lambda_{1,t}+\lambda_{2,t})\theta$, to obtain,
using the usual \dgb\ techniques,
$$\xi_{xx}+\lambda_{1,t}-\xi_x \lambda_1=0.$$  This can be integrated;
substituting
into the equations the solution, along with
$\theta=\exp\{x\lambda_1(t)+\lambda_2(t)\}$,
and reading off coefficients in the independent functions, the
exponentials
in $x$, $2x$, and so on, yields various equations which lead to an
inconsistency.
Thus we need only consider the cases $\theta=0$ and $\theta_x=0$, i.e.,
$\phi_u=0$ and
$\phi_{xu}=0$.

\smallskip\noindent{\sl Subcase 3.2.1 $\phi_u=0$}.
Substituting $\phi_u=0$ into the determining equations we obtain the
following \dgb\ for the system:
$$\eqalignno{
&\xi_u=0,\qquad
\xi_x=0,\qquad
\phi_u=0,&\eqnn{}\cr
&
\beta\phi\phi_{xx}-\phi_{xx}-\beta\phi_{x}^2-\xi\phi_{xxxx}+\xi\phi_{xx}=0,&\eqnn{phiieqn}\cr
& \xi_t(1-\beta\phi)+\beta\xi^2\phi_x+\beta\xi\phi_t=0.&\eqnn{EQvi}
\cr}$$
Thus $\xi$ is an arbitrary function of $t$ and so, as in Case
3.1.2(ii), we set $\xi(t)=\d f/\d t$.
We integrate ({\EQvi}) using the method of characteristics to obtain
$$ \phi=\ft\eta(\zeta) +{1\over\beta},\qquad
\zeta=x-f(t).\eqn{wiieqn}$$
Substituting ({\wiieqn}) into ({\phiieqn}) yields
$$\beta\left[\eta{\d^2\eta\over\d\zeta^2}-\left({\d\eta\over\d\zeta}\right)^2\right]-
{\d^4\eta\over\d\zeta^4}+{\d^2\eta\over\d\zeta^2}=0,\eqn{wiiieqn}$$
which is not of \p\ type. Hence we obtain the infinitesimals
$$\xi = \ft,\qquad\phi=\ft\eta(\zeta) +{1\over\beta},\eqn{}$$
where $\zeta=x-f(t)$ and $\eta(\zeta)$ is a solution of (\wiiieqn).
These yield the (classical) symmetry reduction (\clsrii) with
$z\equiv\zeta=x-f(t)$.

\smallskip\noindent{\sl Subcase 3.2.2 $\phi_{ux}=0$}. In this case we
obtain the solution
$$\xi={[x+\cc6]f(t)},\qquad
\phi=-{f(t)}\left[u+(2x+t+\cc7)/\beta\right]+{1/\beta}, $$
where $f(t)$ is an arbitrary function and $\cc6$ an $\cc7$ are
arbitrary constants.
This is the same as the general case, with $\alpha=-\beta$.

\section{The integrability of the shallow water wave equation
(\eqgsww)}
In this section,  we give first the \p\ analysis of
(\eqgsww), and use the singular manifold method to find
another family of solutions similar, but not equivalent, to
the family (\swwsol), for the special case $\alpha=\beta$.
We show that both (\eqswwi) and (\eqswwii) satisfy the
\p\ property and are solvable by the inverse scattering method,
 suggesting that the solutions found using the
nonclassical and singular manifold methods do not arise
because there exists
some transformation that linearises the equation (\eqswwi).
In fact, it can be seen that the arbitrary function $f(t)$ that
occurs in the families of solutions found arises naturally during
the inverse scattering method of solution.

\subsection{The Painlev\'{e} Tests} We apply the Painlev\'e PDE test
due to Weiss
\etal \cite{\refWTC} to the {\sc gsww} equation ({\eqgsww}).
The {\it \p\ Conjecture\/} (or \p\ ODE test) as formulated by Ablowitz
\etal
\cite{\refARSa,\refARSb} asserts that every \ode\ which arises as a
symmetry reduction of a
completely integrable nonlinear \pde\ is of \p\ type, though perhaps
only after a transformation of
variables. Ablowitz \etal \cite{\refARSb} and McLeod and Olver
\cite{\refMcLO} have given proofs of the \p\ ODE test under certain
restrictions.

Subsequently, Weiss \etal \cite{\refWTC} proposed the \p\ PDE test as a
method of applying the \p\
ODE test directly to a given \pde\ without having to consider symmetry
reductions (which might not
exist). As for the \p\ ODE test, at present there is no rigorous proof
of the \p\ PDE
test, though a partial proof can be inferred from the partial proof of
the \p\ ODE test due to
McLeod and Olver \cite{\refMcLO}. Despite being no means foolproof
the \p\ tests appear to provide a useful criterion for the
identification of completely integrable
\pdes. In addition to providing a valuable first test for whether a
given \pde\ is completely
integrable, other important information can be obtained by use of
\p\ analysis such as B\"acklund
transformations, Lax pairs, Hirota's bilinear representation, special
and rational solutions for
completely integrable equations and special and rational solutions for
nonintegrable equations (see,
for example, \cite{\refNTZ,\refJWx} and the references therein).

To apply the Painlev\'e PDE test to the {\sc gsww} equation ({\eqgsww})
we seek a solution
in the form
$$ u(x,t)=\sum_{k=0}^{\infty}u_k(t)\phi^{k+p}(x,t),\qquad
\phi=x+\psi(t),\eqn{eqIVi}
$$ where $\psi(t)$ is an arbitrary analytic function and $u_k(t)$,
$k=0,1,2,\ldots,$ analytic
functions such that $u_0\not\equiv0$, in the neighbourhood of an
arbitrary, non-characteristic
movable singularity manifold defined by $\phi(x,t)=0$, and $p$ is a
constant to be determined.  By
leading order analysis we find that $p=-1$ and
$u_0(t)=12/(\alpha+\beta)$, provided that
$\alpha+\beta\not=0$. In the case when $\alpha=-\beta$ it is routine to
show that ({\eqgsww}) is
non-\p. We now substitute ({\eqIVi}) into ({\eqgsww}) to obtain from
the
coefficient of $\phi^{k-4}$,
$$
(k+1)(k-1)(k-4)(k-6)u_k=H_k\left(u_{k-1},u_{k-2},\dots,u_0,\psi\right),\eqno{\eqnm{eqIViii}{a}}
$$  where
$$\eqalignno{ H_k &=  (k-3)(k-4)\left(1+{\d\psi\over\d t}\right)u_{k-2}
+ (k-4){\d u_{k-3}\over\d t}
-(k-2)(k-3)(k-4){\d u_{k-1}\over\d t} \cr
&\qquad
-\sum_{j=1}^{k-1}(k-j-1)(j-1)\left[(j-2)\alpha+(k-j-2)\beta\right]u_ju_{k-j}{\d\psi\over\d
t}\cr
&\qquad
-\sum_{j=1}^{k}(k-j-1)\left[(j-2)\alpha+(k-j-2)\beta\right]u_{k-j}{\d
u_{j-1}\over\d t}
&\eqnr{b}
\cr}$$ for $k=0,1,2,\dots$ and where we define $u_k=0$ for $k<0$. This
defines $u_k$ unless
$k=1$, $k=4$ or $k=6$ which are the so-called resonances. At each
positive resonance there is a
compatibility condition which must be identically satisfied for the
expansion ({\eqIViii}) to be
valid, i.e., we require that $H_1\equiv0$, $H_4\equiv0$ and
$H_6\equiv0$ for ({\eqgsww}) to have a
solution of the form ({\eqIViii}).  The compatibility condition
$H_1\equiv0$ is
identically satisfied which implies that $u_1(t)$ is arbitrary.
Equations ({\eqIViii}) with
$k=2$ and $k=3$ yield
$$ u_2={1\over(\alpha+\beta)}\left\{{\d\psi\over\d t}+1 - \beta {\d
v_{1}\over\d t}\right\}
\left({\d\psi\over\d t}\right)^{-1} $$
and
$$  u_3={(\alpha-2\beta)\over2(\alpha+\beta)^2}\left\{\left(\beta
{\d v_{1}\over\d t}-1\right){\d^2\psi\over\d t^2} - \beta
{\d\psi\over\d t}{\d^2 v_{1}\over\d t^2}\right\}\left({\d\psi\over\d
t}\right)^{-3}$$
respectively. The compatibility condition
$H_4\equiv0$ yields
$$\eqalignno{
&{12(\alpha-\beta)(\alpha-2\beta)\over(\alpha+\beta)^3}\left\{\left(\beta
{\d
v_{1}\over\d t}-1\right)\left[{\d\psi\over\d t}{\d^3\psi\over\d
t^3}-3\left({\d^2\psi\over\d
t^2}\right)^2\right]\right.\cr &\qquad\left.+2\beta {\d^2 v_{1}\over\d
t^2}{\d\psi\over\d
t}{\d^2\psi\over\d t^2}-\beta {\d^3 v_{1}\over\d
t^3}\left({\d\psi\over\d t}\right)^2
\right\}=0.\cr}$$   Since $\psi$ is an arbitrary function this implies
that
$$(\alpha-\beta)(\alpha-2\beta)=0,\eqn{eqIvv}$$ is a necessary
condition for ({\eqgsww}) to
have a solution of the form ({\eqIViii}). The compatibility condition
$H_6\equiv0$ is also satisfied if and only if ({\eqIvv}) is satisfied.
Therefore we conclude that
({\eqgsww}) has a solution in the form ({\eqIVi}) if either (i),
$\alpha=\beta$ or (ii),
$\alpha=2\beta$. These are the same conditions for ({\clsriic}), which
arises in the classical
reduction, to be of \p-type and for the {\sc gsww} equation ({\eqgsww})
to expressible in Hirota's
bilinear form. If
$(\alpha-\beta)(\alpha-2\beta)\not=0$, then it is necessary to
introduce a
$v_4(t)\phi^3(x,t)\ln\phi(x,t)$ term, where $v_4(t)$ is to be
determined, into the expansion
({\eqIVi}) and at higher orders of
$\phi(x,t)$, higher and higher powers of $\ln\phi(x,t)$ are required; a
strong indication
of non-\p\ behaviour. Hence the \p\ PDE test suggests that
$\alpha=\beta$ and $\alpha=2\beta$ are
the only integrable cases of the {\sc gsww} equation ({\eqgsww}).

Exact solutions of the {\sc swwi} equation (\eqswwi) can be obtained
using the so-called
singularity manifold method which uses truncated \p\ expansions
\cite{\refWTC,\refWeiss}. If we seek
a solution of (\eqswwi) in the form $$u(x,t) =
{6\over\beta}\,{\phi_x(x,t)\over\phi(x,t)},\eqn{}$$
and then equate coefficients of powers of $\phi$ to zero, we find that
$\phi(x,t)$ satisfies the
overdetermined system
$$\eqalignno{&\phi_{xxxt} - \phi_{xx} - \phi_{xt}
=0,&\eqnm{smeqs}{a}\cr &\phi_t\phi_{xxx} -
3\phi_{xt}\phi_{xx} - \phi_{x}^2 + \phi_x\left(3\phi_{xxt}
- \phi_{t}\right) =0.&\eqnr{b}\cr}$$
(A {\sc dgb} analysis of this system leads to some very
complex expressions. Although it does yield some \odes\ in $x$ for
$\phi$ in the various subcases they appear difficult to solve.)  Now
suppose we seek a solution of these equations in the form
$$\phi(x,t) = a_1\exp\left\{\kappa_1x+\mu_1t\right\}+
a_2\exp\left\{\kappa_2x+\mu_2t\right\} +
a_0,\eqn{smsol}$$ where $a_0$, $a_1$, $a_2$, $\kappa_1$, $\kappa_2$,
$\mu_1$ and $\mu_2$ are
constants. It is straightforward to show that equations (\smeqs) have a
solution of the form (\smsol)
provided that
$\mu_1 = \kappa_1/(\kappa_1^2-1)$, $\mu_2 = \kappa_2/(\kappa_2^2-1)$
and $\kappa_1$ and
$\kappa_2$ satisfy the constraint
$$\kappa_1^2-\kappa_1\kappa_2+\kappa_2^2=3.\eqn{smcond}$$ Thus we
obtain the following exact
solution of the {\sc swwi} equation (\eqswwi) given by
$$u(x,t)
={\displaystyle6\left[a_1\kappa_1\exp\left\{\kappa_1x+{\kappa_1t\over\kappa_1^2-1}\right\}+
a_2\kappa_2\exp\left\{\kappa_2x+{\kappa_2t\over\kappa_2^2-1}\right\}\right]
\over \displaystyle
\beta\left[a_1\exp\left\{\kappa_1x+{\kappa_1t\over\kappa_1^2-1}\right\}+
a_2\exp\left\{\kappa_2x+{\kappa_2t\over\kappa_2^2-1}\right\} +
a_0\right]},\eqn{swwsmsol}$$ provided
$\kappa_1$ and $\kappa_2$ satisfy (\smcond).

It should be noted that (\swwsmsol) and (\swwsolii) are fundamentally
different solutions of
the {\sc swwi} equation (\eqswwi) as we shall now demonstrate. The
general two-soliton solution of
(\eqswwi) is given by
$$u(x,t) =
{6\over\beta}\,{1+\kappa_1\exp\left(\eta_1\right)+\kappa_2\exp\left(\eta_2\right)
+A_{12}(\kappa_1+\kappa_2)\exp\left(\eta_1+\eta_2\right)\over
1+\exp\left(\eta_1\right)+\exp\left(\eta_2\right)
+A_{12}\exp\left(\eta_1+\eta_2\right)},\eqno\eqnm{swwiisol}{a}$$ where
$$ \eta_j = \kappa_jx+{\kappa_jt\over\kappa_j^2-1}+\delta_j,\quad
j=1,2,\qquad
A_{12} =
{(\kappa_1-\kappa_2)^2(\kappa_1^2-\kappa_1\kappa_2+\kappa_2^2-3)\over
(\kappa_1+\kappa_2)^2(\kappa_1^2+\kappa_1\kappa_2+\kappa_2^2-3)},\eqno\eqnr{b}$$
with $\kappa_1$,
$\kappa_2$, $\delta_1$ and $\delta_2$ arbitrary constants \cite{\refHS}
(see also \cite{\refMLD}).
The solution (\swwsmsol) is the special case of (\swwiisol) with
$A_{12}=0$; Hirota and Ito
\cite{\refHI} refer to it as being  the ``resonant state'' where either
two solitons fuse together
after colliding with each other or a single soliton splits into two
solitons (see Figure 3a). On the
other hand, (\swwsolii) is the special case of (\swwiisol) with
$A_{12}=1$ where two solitons pass
through each other with no phase shift as a consequence of the
interaction (see Figure 3b). Thus
whereas both solutions are asymptotically equivalent as $t\to\infty$,
they are qualitatively very
different as $t\to-\infty$. This shows that nonclassical method and the
singular manifold method do
not, in general, yield the same solution set.

\subsection{Inverse Scattering}
The inverse scattering method, originally developed by Gardner \etal
\cite{\refGGKM} in order to
solve the {\sc kdv} equation (\eqkdv), has led to the solution of
numerous physically significant
nonlinear evolution equations, such as the nonlinear Schr\"odinger,
Sine-Gordon, Modified {\sc kdv}
and Boussinesq equations (cf.\ \cite{\refAC}). Nonlinear evolution
equations solvable
by inverse scattering are known to possess a number of remarkable
properties which
appear to characterise the equations, including: the existence of
multi-soliton solutions, an
infinite number of symmetries and conservation laws, B\"acklund
transformations, a Lax pair, a
bi-Hamiltonian representation, a prolongation structure, the Hirota
bilinear representation, and the
Painlev\'e  property (cf.\ \cite{\refAC}). However, the precise
relationship between
these properties has yet to be rigorously established.

There are two special cases of the {\sc gsww} equation (\eqgsww) which
have been studied from the
inverse scattering point of view, namely the {\sc swwi} equation
(\eqswwi) and
the {\sc swwii} equation (\eqswwii), or equivalently (\eqswwa) and
(\eqswwb), respectively. Hirota
and Satsuma \cite{\refHS} studied both ({\eqswwa}) and ({\eqswwb})
using Hirota's bilinear technique
\cite{\refHirt}. Equation ({\eqswwb}) is known to be solvable by
inverse scattering \cite{\refAKNS}.
Several of the aforementioned properties of completely integrable
equations
have been derived for ({\eqswwi},{\eqswwii})
\cite{\refMLD,\refHI,\refCM--\refJWv}.

The scattering problem for the {\sc swwii} equation (\eqswwii) is the
second order problem
\cite{\refAKNS} $$\psi_{xx} + \tfr12\beta u_x \psi =
\lambda\psi,\eqn{scpiia}$$
with associated time-dependence
$$(4\lambda-1)\psi_t = (1- \beta u_t)\psi_x  +\tfr12\beta
u_{xt}\psi,\eqn{scpiib}$$ where $\lambda$
is the constant eigenvalue, and $\psi_{xxt}=\psi_{txx}$ if and only if
$u$ satisfies (\eqswwii) We note that (\scpiia) is the time-independent
Schr\"odinger equation
which is also the scattering
problem for the {\sc kdv} equation (\eqkdv) \cite{\refGGKM}. In
contrast, the scattering problem for
the {\sc swwi} equation (\eqswwi) is the third order problem
\cite{\refCM,\refMus}
$$\psi_{xxx} + \left(\tfr12\beta u_x - 1\right)\psi_x =
\lambda\psi,\eqn{scpia}$$
with associated time-dependence
$$3\lambda\psi_t = (1- \beta u_t)\psi_{xx}  + \beta
u_{xt}\psi_x.\eqn{scpib}$$
We remark that (\scpia) is similar to the scattering problem
$$\psi_{xxx} + \tfr14(1+6u)\psi_x +\tfr34\left[u_x -
\i\sqrt{3}\,\partial_x^{-1}(v_t)\right]\psi=
\lambda\psi\eqn{bqspa}$$  which is the scattering problem for the
Boussinesq equation
$$u_{xxxx} + 3(u^2)_{xx} + u_{xx} = u_{tt},\eqn{eqbq}$$
and which has been comprehensively studied by Deift \etal
\cite{\refDTT}.

Only the derivative $u_x$ appears in the scattering problem (\scpia)
and so an arbitrary function
of $t$ may be added to $u$ without affecting this. This function can be
fixed by the requirement that
$u_t(x,t)\to 0$ as $x\to\infty$; the scattering problem (\scpia) is
solvable for
$u$ such that $u_t(x,t)\to 0$ sufficiently rapidly as $x\to\infty$.
Moreover, the associated
time-dependence, (\scpib), is invariant under the variable-coefficient
Galilean transformation
(\eqGalTr). Furthermore, note that one can integrate the {\sc swwi}
equation (\eqswwi) once with
respect to $x$ and so introduce an arbitrary function of $t$.

\section{Discussion}
In this paper we have discussed the shallow water equation (\eqgsww).
In particular, for the special
case of (\eqgsww) given by the {\sc swwi} equation (\eqswwi), using the
nonclassical symmetry
reduction method  originally proposed by Bluman and Cole
\cite{\refBCa}, we obtained a family of
solutions (\swwsol) which have a rich variety of qualitative
behaviours. This is due to the freedom
in the choice of the arbitrary function $f(t)$. One can choose $f_1(t)$
and $f_2(t)$ such
$|f_1(t)-f_2(t)|$ is exponentially small as $t\to\infty$, yet $f_1(t)$
and $f_2(t)$ are quite
different as $t\to-\infty$, so that as $t\to\infty$ the two solutions
are essentially the same, yet
as $t\to-\infty$ they are radically different. In Figure 1 we show that
by a judicious choice of
$f(t)$ we can exhibit a plethora of different solutions.

We believe that these results suggest that solving the {\sc swwi}
equation (\eqswwi) numerically for
initial conditions such as those in the solutions plotted in Figure 1
could pose some fundamental
difficulties. An exponentially small change in the initial data yields
a fundamentally different
solution as $t\to-\infty$. How can any numerical scheme in current use
cope with such behaviour?
Recently Ablowitz \etal \cite{\refASH} have shown that the focusing
nonlinear \sch\ equation
$$\i u_t + u_{xx} + |u|^2u=0,\eqn{eqnls}$$ exhibits numerical chaos
created by small errors on the
order of roundoff. The results of Ablowitz \etal together with those
given in this paper
suggest that numerical analysts need to take care to ensure the
accuracy of their programs.

The solution (\swwsol) appears to be a nonlinear superposition of
solutions suggesting that the {\sc
swwi} equation (\eqswwi) may be linearisable through a transformation
to a linear \pde, analogous to
the linearisation of Burgers' equation
$$ u_t  =  u_{xx} + 2uu_x, \eqn{burger} $$ which is mapped to the
linear heat equation through the
Cole-Hopf transformation \cite{\refCole,\refHopf}. If so then the
solution (\swwsol) could be viewed as an artefact of the fact that the
{\sc swwi} equation (\eqswwi)
is linearisable. However as illustrated in \S4, the {\sc swwi} equation
(\eqswwi) can be expressed as
the compatibility condition of a third order spectral problem. Further
the associated scattering
problem (\scpia) is very similar to that for the Boussinesq equation
which has been thoroughly
studied by Deift \etal  \cite{\refDTT}. This strongly suggests the {\sc
swwi} equation (\eqswwi) is
solvable by inverse scattering. Additionally, as mentioned in \S4, the
spatial part of the inverse
scattering formalism (\scpia) only defines $u$ up to an arbitrary
additive function of $t$;
this arbitrary function may be incorporated into $u$ using the
variable-coefficient Galilean
transformation (\eqGalTr).

Since the generalised shallow water equation (\eqgsww) is invariant
under the variable-coefficient
Galilean transformation (\eqGalTr) for all $\alpha$ and $\beta$, one
can take any solution of the
equation and using (\eqGalTr) generate some interesting solutions.

Fujioka and Espinosa \cite{\refFE} have discussed symmetry reductions
of (\eqswwa) using the
classical Lie method and direct method due to Clarkson and Kruskal
\cite{\refCK}. They claim that the
classical method yields no symmetry reductions and that the direct
method yields symmetry reductions
that are a subset of those we obtained in \S2 using the classical
method. When we applied the
nonclassical method to  (\eqswwi) in \S3, we found that $\xi_u=0$,
consequently the results of Olver
\cite{\refOlverb} (see also \cite{\refABH,\refPucci}) show that the
direct and nonclassical
methods yield the same reductions. The difficulty Fujioka and Espinosa
\cite{\refFE} appear to have
experienced is with the nonlocal term in (\eqswwa); considering
(\eqswwi) rather than (\eqswwa)
seems to  be simpler.

\appendix{The solution of (\clsriic)}
\def\sen{\hbox{A}}
In this appendix we show how equation (\clsriic), in the special case
when $\alpha=\beta$, i.e.,
$$z\wzzzz+4\wzzz+ 2\beta z\wz\wzz + \beta w\wzz +
2\beta\left(\wz\right)^2=0, \eqn{appia}$$ can be
solved in terms of the third \p\ equation (PIII) \cite{\refInce}
$${\d^2y\over\d x^2} = {1\over y}\left({\d y\over\d x}\right)^2-
{1\over x}{\d y\over\d x} + a y^3 +
{b y^2+c\over x} + {d\over y},\eqn{eqpiii}$$ with $a$, $b$, $c$ and $d$
constants, and also in terms
of the fifth \p\ equation (PV)
$${\d^2y\over\d x^2} = \left\{{1\over 2y}+{1\over y-1}\right\}\left({\d
y\over\d x}\right)^2-
{1\over x}{\d y\over\d x}+ {(y-1)^2\over x^2}\left\{a y + {b\over
y}\right\} + {cy\over x} +
{dy(y+1)\over y-1},\eqn{eqpv}$$ with $a$, $b$, $c$ and $d$ constants.
We also demonstrate how in the
special case when $\alpha=2\beta$, i.e.,
$$z\wzzzz+4\wzzz+ 3\beta z\wz\wzz + \beta w\wzz +
4\beta\left(\wz\right)^2=0, \eqn{appib}$$
(\clsriic) can be solved in terms of PIII.

Cosgrove and Scoufis \cite{\refCS} consider the the equation
$$\eqalignno{X^2\left({\d^2Y\over\d X^2}\right)^2 = -4&\left({\d
Y\over\d X}\right)^2
\left(X{\d Y\over\d X} - Y\right) + A_1\left(X{\d Y\over\d X} -
Y\right)^2
\cr&+ A_2\left(X{\d Y\over\d X} - Y\right) + A_3{\d Y\over\d X} +
A_4,&\eqnn{eqcsvv}\cr}$$ where
$A_1$, $A_2$, $A_3$ and $A_4$ are constants (their equation SD-I.b
(5.5)).
In the general case
when $A_1$ and $A_2$ are not both zero, Cosgrove and Scoufis show that
(\eqcsvv) is solvable in
terms of PV (\eqpv) though the transformation
$$\eqalignno{ Y(X) &= {1\over4y}\left({x\over y-1}\yp - y\right)^2 -
\tfr14(1-\sqrt{2}a)^2(y-1) -
\tfr12b{(y-1)\over y} \cr &\qquad+ \tfr14cx{y+1\over y-1} + \tfr12 d
{x^2y\over(y-1)^2},&\eqnm{transa}{a}\cr  X &= x,&\eqnr{b}\cr }$$ where
$$\eqalignno{A_1 = -2d,&\qquad  A_2 = \tfr14c^2 + 2bd -
d(1-\sqrt{2}a)^2,&\eqnm{transai}{a}\cr
A_3 = bc + \tfr12c(1-\sqrt{2}a)^2,&\qquad A_4 =
\tfr18c^2\left[(1-\sqrt{2}a)^2-2b\right] -
\tfr18d\left[(1-\sqrt{2}a)^2+2b\right]^2.{\hbox to
20pt{\hfill}}&\eqnr{b}\cr}$$  In the case when
$A_1=0$ and $A_2$ is unrestricted, Cosgrove and Scoufis show that
(\transa) is solvable in terms of
PIII (\eqpiii) through the transformation
$$Y(X) = {1\over16y^2}\left({x}\yp - y\right)^2 -\tfr1{16}ax^2y^2
-\tfr18\left(b+2\sqrt{a}\right){x}y + {cx\over8y} +
{dx^2\over16y^2},\qquad X = x^2,\eqn{transb}$$
where
$$A_2 = -\tfr1{16}ad,\qquad A_3 =
\tfr1{16}c\left(c+2\sqrt{a}\right),\qquad   A_4 =
{1\over256}\left[ac^2 -
d\left(c+2\sqrt{a}\right)^2\right].\eqn{transbi}$$  Therefore if
$A_1=0$ and
$A_2\not=0$ then (\transa) is solvable in terms of both PIII (\eqpiii)
and PV (\eqpv) since the
special case of PV with $d=0$ can always be solved in terms of
solutions of  PIII
\cite{\refFA,\refGromak}; Cosgrove and Scoufis \cite{\refCS} remark
that there are infinitely many
other special cases of PV and (\transa) that are solvable in terms of
solutions of PIII, e.g., if
$A_3=0$ and $A_2^2 = 4A_1A_4$ in (\transa).

To illustrate how (\appia) and (\appib) are solvable in terms of PIII,
we differentiate (\eqcsvv) with respect to $X$ to yield
$$X{\d^3Y\over\d X^3}+ {\d^2Y\over\d X^2} = -6\left({\d Y\over\d
X}\right)^2 +  {4Y\over X}{\d
Y\over\d X} + A_1\left(X{\d Y\over\d X} - Y\right) + \tfr12A_2 +
{A_3\over2 X}.\eqn{eqcsvvb}$$
Integrating (\appia) once yields
$$z\wzzz+3\wzz+ 2\beta z\left(\wz\right)^2 + \beta w\wz = B_1,
\eqn{appiia}$$ with $B_1$ an
arbitrary constant. Now making the transformation
$$ w(z) = {Y(X)\over z} - {1\over 4\beta z},\qquad X=z^{3/2},$$ and
setting $\beta=9$ (without loss
of generality) yields (\eqcsvvb) with $A_1=0$, $A_2=-16B_1/27$ and
$A_3=0$. Therefore (\appiia) is
solvable in terms of PV with
$$a=\tfr12\left[1-\tfr34\sqr{-3A_4\over B_1}\right]^2,\qquad
b=-{27A_4\over32B_1},\qquad c=
\sqr{-64B_1\over27},\qquad d=0,$$ and in terms of PIII with either (i)
$a$ and $b$ arbitrary,
$c=0$ and $d=256B_1/(27a)$, or (ii), $a$ and $c$ arbitrary,
$b=-2\sqrt{a}$ and $d=256B_1/(27a)$.

Analogously integrating (\appib) once yields
$$z^2\wzzz+2z\wzz -2\wz+ \tfr32\beta z\left(\wz\right)^2 + \beta
\left(zw\wz -\tfr12 w^2\right)=B_2,
\eqn{appiib}$$ with $B_2$ an arbitrary constant. Then making the
transformation
$$ w(z) = {Y(X)\over z} - {1\over 2\beta z},\qquad X=z^{2},$$ and
setting $\beta=8$ yields
(\eqcsvvb) with $A_1=0$, $A_2=0$ and $A_3=-\tfr14 B_2$. Thus (\appiib)
is solvable in terms of PIII
with either (i) $c$ and $d$ arbitrary, $a=0$ and $b=-4B_2/c$, or (ii)
$a$ and $b$ arbitrary,
$c=-4B_2/(b+2\sqrt{a})$ and $d=0$.

We remark that (\eqcsvv), or equations that are equivalent to (\eqcsvv)
through a Lie point
transformation, appear in the work of Bureau
\cite{\refBurea,\refBureb}, Chazy \cite{\refChazy},
Cosgrove \cite{\refCosa,\refCosb}, Jimbo \cite{\refJimbo} and Jimbo and
Miwa \cite{\refJMii} (see
\cite{\refCS} for further details).

It is well known that PIII (\eqpiii) possesses rational solutions and
one-parameter families of
solutions expressible in terms of Bessel functions (cf.,
\cite{\refFA,\refAir--\refOka}) and
B\"acklund transformations which map solutions of PIII into new
solutions for PIII but for different
values of the  parameters (cf.,
\cite{\refFA,\refGromak,\refAir--\refLukb}). Starting with these
known rational and one-parameter family solutions, hierarchies of
solutions of PIII can be generated
by means of the above B\"acklund transformations (cf.,
\cite{\refMilne}). Analogously, it is well
known that PV (\eqpv) possesses rational solutions and one-parameter
families of solutions
expressible in terms of Whittaker functions (cf.,
\cite{\refFA,\refAir--\refLuka,\refKLM,\refOkb})
and B\"acklund transformations which map solutions of the equation into
new solutions with different
values of the parameters (cf.,
\cite{\refFA,\refAir,\refGrob,\refMFokb,\refOkb--\refLukc}). Using
these special exact solutions of PIII (\eqpiii) and PV (\eqpv), one can
construct exact solutions of
(\eqswwi) and (\eqswwii), though we do pursue this further here.

\ack
It is a pleasure to thank Mark Ablowitz and Jim Curry for several
illuminating discussions. We also
thank the Program in Applied Mathematics, University of Colorado at
Boulder  and the Department of
Mathematics and Statistics, University of Pittsburgh for their
hospitality during our visits and
Chris Cosgrove and Willy Hereman for expert suggestions and comments.
The support of SERC (grant
GR/H39420) is gratefully acknowledged.  PAC is also grateful for
support through a Nuffield
Foundation Science Fellowship and NATO grant CRG 910729.

\def\refpp#1#2#3{{\rm#1}\ #2, #3}
\def\refjl#1#2#3#4#5{{\rm#1}\  #2, {\frenchspacing\it#3}\ {\bf#4}\ #5}
\def\refbk#1#2#3#4{{\rm#1}\  #2, ``{\sl#3}'' #4.}
\def\refcf#1#2#3#4#5#6{{\rm#1}\  #2, in ``{\sl#3}'' [#4]\ #5\ #6}

\def\fit{\frenchspacing\it}
\def\refn#1{\item{[\expandafter \csname #1\endcsname]\ }}

\references
\parindent=30pt

\refn{refWhitham}
\refbk{Whitham G B}{1974}{Linear and Nonlinear Waves}{Wiley, New York}

\refn{refHS}
\refjl{Hirota R and Satsuma J}{1976}{J. Phys. Soc. Japan}{40}{611--612}
\refn{refAKNS}
\refjl{Ablowitz M J, Kaup D J, Newell A C and Segur H}{1974}{Stud.
Appl. Math.}{53}{249--315}
\refn{refGGKM}
\refjl{Gardner C S, Greene J M, Kruskal M D and Miura R M}{1967}{Phys.
Rev. Lett}{19}{1095--1097}
\refn{refPer}
\refjl{Peregrine H}{1966}{J. Fluid Mech.}{25}{321--330}
\refn{refBBM}
\refjl{Benjamin T B, Bona J L and Mahoney J}{1972}{Phil. Trans. R. Soc.
Lond. Ser. A}{272}{47--78}
\refn{refMcLO}
\refjl{McLeod J B and Olver P J}{1983}{SIAM
J.\ Math.\ Anal.}{14}{488--506}

\refn{refHiet}
\refcf{Hietarinta J}{1990}{Partially Integrable Evolution Equations in
Physics}{Eds.\ R.\ Conte and
N.\ Boccara}{{\it NATO ASI Series C: Mathematical and Physical
Sciences\/}, {\bf 310}, Kluwer,
Dordrecht}{pp459--478}
\refn{refHirt}
\refcf{Hirota R}{1980}{Solitons}{Eds.\ R.K.\ Bullough and
P.J.\ Caudrey}{{\it Topics in Current
Physics\/}, {\bf 17}, Springer-Verlag, Berlin}{pp157--176}

\refn{refARSa}
\refjl{Ablowitz M J, Ramani A and Segur H}{1978}{\PRL}{23}{333--338}
\refn{refARSb}
\refjl{Ablowitz M J, Ramani A and Segur H}{1980}{\jmp}{21}{715--721}
\refn{refWTC}
\refjl{Weiss J, Tabor M and Carnevale G}{1983}{\jmp}{24}{522--526}

\refn{refBLMP}
\refjl{Boiti M, Leon J J-P, Manna M and Pempinelli
F}{1986}{\IP}{2}{271--279}
\refn{refJM}
\refjl{Jimbo M and Miwa T}{1983}{Publ. R.I.M.S.}{19}{943--1001}
\refn{refDGRW}
\refjl{Dorizzi B, Grammaticos B, Ramani A and Winternitz
P}{1986}{\JMP}{27}{2848--2852}

\refn{refYOS}
\refjl{Yajima N, Oikawa M and Satsuma J}{1978}{\JPSJ}{44}{1711--1714}
\refn{refKY}
\refjl{Kako F and Yajima N}{1980}{\JPSJ}{49}{2063--2071}

\refn{refBogi}
\refjl{Bogoyaviemskii O I}{1990}{Math. USSR Izves.}{34}{245--259}
\refn{refBogii}
\refjl{Bogoyaviemskii O I}{1990}{Russ. Math. Surv.}{45}{1--86}

\refn{refBK}
\refbk{Bluman G W and Kumei S}{1989}{Symmetries and Differential
Equations}{{\fit Appl. Math. Sci.\/}, {\bf 81}, Springer-Verlag,
Berlin}
\refn{refOlver}
\refbk{Olver P J}{1993}{Applications of Lie Groups to Differential
Equations}{2nd Edition, Springer Verlag, New York}
\refn{refHere}
\refjl{Hereman W}{1993}{Euromath Bull.}{2}{to appear}
\refn{refCHW}
\refjl{Champagne B, Hereman W and Winternitz P}{1991}{Comp. Phys.
Comm.}{66}{319--340}
\refn{refBCa}
\refjl{Bluman G W and Cole J D}{1969}{J.\ Math.\ Mech.}{18}{1025--1042}

\refn{refLW}
 \refjl{Levi D and Winternitz P}{1989}{\jpa}{22}{2915--2924}

\refn{refVor}
\refjl{Vorob'ev E M}{1991}{Acta Appl. Math.}{24}{1--24}

 \refn{refCK}
\refjl{Clarkson P A and Kruskal M D}{1989}{\jmp}{30}{2201--2213}

 \refn{refPACrev}
 \refjl{Clarkson P A}{1993}{Math. Comp. Model.}{18}{45--68}
 \refn{refFush}
\refjl{Fushchich W I}{1991}{Ukrain. Mat. Zh.}{43}{1456--1470}

 \refn{refGalaka}
\refjl{Galaktionov V A}{1990}{Diff. and Int. Eqns.}{3}{863--874}
 \refn{refGDEKS}
 \refjl{Galaktionov V A, Dorodnytzin V A, Elenin G G, Kurdjumov S P and
 Samarskii A A}{1988}{J. Sov. Math.}{41}{1222--1292}
 \refn{refAmesii}
 \refjl{Ames W F}{1992}{Appl. Num. Math.}{10}{235--259}
 \refn{refShok}
 \refbk{Shokin Yu I}{1983}{The Method of Differential
 Approximation}{Springer-Verlag, New
York}

 \refn{refReida}
\refjl{Reid G J}{1990}{\jpa}{23}{L853--L859}
 \refn{refReidb}
\refjl{Reid G J}{1991}{Europ. J. Appl. Math.}{2}{293--318}
\refn{refMF}
\refpp{Mansfield E and Fackerell E}{1992}{``Differential Gr\"obner
Bases",
preprint {\bf 92/108}, Macquarie University, Sydney, Australia}
\refn{refCMa}
\refjl{Clarkson P A and Mansfield E L}{1994}{Physica D}{70}{250--288}
\refn{refZwil}
\refbk{Zwillinger D}{1992}{Handbook of Differential Equations}{Second
Edition, Academic,  Boston}

 \refn{refCMb}
\refjl{Clarkson P A and Mansfield E L}{1994}{SIAM J. Appl. Math.}{}{to
appear}
\refn{refSchw}
\refjl{Schwarz F}{1992}{Computing}{49}{95--115}
\refn{refTop}
\refjl{Topunov V L}{1989}{Acta Appl. Math.}{16}{191--206}
\refn{refBuchi}
\refcf{Buchberger B}{1988}{Mathematical Aspects of Scientific
Software}{\rm Ed.\ J.\ Rice}{Springer
Verlag}{pp59--87}

\refn{refPank}
\refjl{Pankrat'ev E V}{1989}{Acta Appl. Math.}{16}{167--189}

\refn{refRW}
\refpp{Reid G J and Wittkopf A}{1993}{``A Differential Algebra Package
for
{\sc maple}'', {\tt ftp 137.82.36.21} login: anonymous, password: your
email address,
directory: {\tt pub/standardform}}

\refn{refMD}
\refpp{Mansfield E}{1993}{``{\tt diffgrob2}: A symbolic algebra package
for analysing systems of PDE
using Maple", {\tt ftp 137.111.216.12},  login: anonymous, password:
your email address,
directory: {\tt pub/maths/Maple}, files:{\tt diffgrob2\_src.tar.Z,
diffgrob2\_man.tex.Z}}

\refn{refWeiss}
\refjl{Weiss J}{1983}{\jmp}{24}{1405--1413}

\refn{refInce}
\refbk{Ince E L}{1956}{Ordinary Differential Equations}{Dover, New
York}
\refn{refWW}
\refbk{Whittaker E E and  Watson G M}{1927}{Modern Analysis}{4th
Edition, C.U.P., Cambridge}
 \refn{refAndIb}
\refbk{Anderson R L and Ibragimov N H}{1979}{Lie-B\"acklund
Transformations in
Applications}{SIAM, Philadelphia}
\refn{refNTZ}
\refjl{Newell A C, Tabor M and Zeng Y B}{1987}{Physica}{29D}{1--68}
\refn{refJWx}
\refcf{Weiss J}{1990}{Solitons in Physics, Mathematics and Nonlinear
Optics}{Eds P.J.\ Olver and
D.H.\ Sattinger}{{\it IMA Series\/}, {\bf 25}, Springer-Verlag,
Berlin}{pp175--202}

\refn{refMLD}
\refjl{Musette M, Lambert F and Decuyper J
C}{1987}{\JPA}{20}{6223--6235}
\refn{refHI}
\refjl{Hirota R and Ito M}{1983}{\JPSJ}{52}{744--748}

\refn{refAC}
\refbk{Ablowitz M J and Clarkson P A}{1991}{Solitons, Nonlinear
Evolution Equations and Inverse
 Scattering}{{\frenchspacing\it L.M.S. Lect. Notes Math.}, {\bf 149},
 C.U.P., Cambridge}
\refn{refCM}
\refjl{Conte R and Musette M}{1991}{\jmp}{32}{1450--1457}
\refn{refHietb}
\refjl{Hietarinta J}{1987}{\jmp}{28}{1732--1742}
\refn{refHL}
\refjl{Hu X B and Li Y}{1983}{\JPA}{24}{1979--1986}
\refn{refMat}
\refjl{Matsuno Y}{1990}{\JPSJ}{59}{3093--3100}
\refn{refMus}
\refcf{Musette M}{1987}{\p\ Transcendents: Their Asymptotics and
Physical Applications}{Eds.\ D.\
Levi and P.\ Winternitz}{{\it NATO ASI Series B: Physics\/}, {\bf
278},  Plenum, New
York}{pp197--209}
\refn{refTag}
\refjl{Tagami Y}{1989}{\pl}{141A}{116--120}
\refn{refJWv}
\refjl{Weiss J}{1985}{\JMP}{26}{2174--2180}

\refn{refDTT}
\refjl{Deift P, Tomei C and Trubowitz E}{1982}{Commun. Pure Appl.
Math.}{35}{567--628}
\refn{refCole}\refjl{Cole J D}{1951}{Quart.\ Appl.\ Math.}{9}{225--236}
\refn{refHopf}\refjl{Hopf E}{1950}{Commun.\ Pure
Appl.\ Math.}{3}{201--250}
\refn{refASH}\refjl{Ablowitz M J, Schober C and Herbst B
M}{1993}{\PRL}{71}{2683--2686}
\refn{refOlverb}
\refpp{Olver P J}{1993}{``Direct reduction and differential
constraints", preprint, Department of
Mathematics, University of Maryland, College Park, MD}
\refn{refABH}
\refjl{Arrigo D J, Broadbridge P and Hill J
M}{1993}{\jmp}{34}{4692--4703}
\refn{refPucci}
\refjl{Pucci E}{1992}{\jpa}{25}{2631--2640}
\refn{refFE}
\refjl{Fujioka J and Espinosa A}{1980}{\JPSJ}{60}{4071--4075}

\refn{refCS}
\refjl{Cosgrove C M and Scoufis G}{1993}{\sam}{88}{25--87}
\refn{refFA}
\refjl{Fokas A S and Ablowitz M J}{1983}{\jmp}{23}{2033--2042}
\refn{refGromak}
\refjl{Gromak V I}{1975}{Diff. Eqns.}{11}{285--287}

\refn{refBurea}
\refjl{Bureau F}{1972}{Ann. Mat. Pura Appl. (IV)}{91}{163--281}
\refn{refBureb}
\refjl{Bureau F, Garcet A and Goffar J}{1972}{Ann. Mat. Pura Appl.
(IV)}{92}{177--191}
\refn{refChazy}
\refjl{Chazy J}{1911}{Acta Math.}{34}{317--385}
\refn{refCosa}
\refjl{Cosgrove C M}{1977}{\jpa}{10}{2093--2105}
\refn{refCosb}
\refjl{Cosgrove C M}{1978}{\jpa}{11}{2405--2430}
\refn{refJimbo}
\refjl{Jimbo M}{1982}{Publ. RIMS, Kyoto Univ.}{18}{1137--1161}
\refn{refJMii}
\refjl{Jimbo M and Miwa T}{1981}{Physica}{D2}{407--488}
\refn{refAir}
\refjl{Airault H}{1979}{Stud. Appl. Math.}{61}{31--53}
\refn{refGrob}
\refjl{Gromak V I}{1978}{Diff. Eqns.}{14}{1510--1513}
\refn{refLuka}
\refjl{Lukashevich N A}{1965}{Diff. Eqns.}{1}{561--564}
\refn{refMFokb}
\refjl{Mugan U and Fokas A S}{1992}{\jmp}{33}{2031--2045}
\refn{refOka}
\refjl{Okamoto K}{1987}{Funkcial. Ekvac.}{30}{305--332}
\refn{refLukb}
\refjl{Lukashevich N A}{1967}{Diff. Eqns.}{3}{994--999}
\refn{refMilne}
\refcf{Milne A E and Clarkson P A}{1993}{Applications of Analytic and
Geometric Methods to Nonlinear
Differential Equations}{Editor P.A.\ Clarkson}{{\it NATO ASI Series C:
Mathematical and Physical
Sciences\/}, Kluwer, Dordrecht}{pp341--352}
\refn{refKLM}
\refjl{Kitaev A V, Law C K and McLeod J B}{1993}{J. Diff. Int.
Eqns.}{}{to appear}
\refn{refOkb}
\refjl{Okamoto K}{1987}{Jap. J. Math.}{13}{47--76}
\refn{refGroc}
\refjl{Gromak V I}{1976}{Diff. Eqns.}{12}{740--742}
\refn{refLukc}
\refjl{Lukashevich N A}{1968}{Diff. Eqns.}{4}{1413--1420}

\vfill\eject

\figures
\noindent{\bf Figure 1.}
\hangindent=16pt\hangafter=1  The solution (\swwsoli) where
\hfill\break{\hbox to 30pt{(i),\hfill}}$f(t)=\tfr12t$,
\hfill\break{\hbox to 30pt{(ii),\hfill}}$f(t)=\tfr12t+\exp(-t/10)$,
\hfill\break{\hbox to 30pt{(iii),\hfill}}$f(t)=\tfr12t+[1-\tanh t]\sin
t$,
\hfill\break{\hbox to 30pt{(iv),\hfill}}$f(t)=\tfr12t+ 2\hbox{Ai}(2t)$,
\hfill\break{\hbox to 30pt{(v),\hfill}}$f(t)=\tfr12t+2\exp(-t^2/20)\sin
t$,
\hfill\break{\hbox to
30pt{(vi),\hfill}}$f(t)=\tfr12t+\exp(-t^2/100)\sin t$,
\hfill\break{\hbox to
30pt{(vii),\hfill}}$f(t)=\tfr12t+2\pi^{-1}\tan^{-1} t$,
\hfill\break{\hbox to 30pt{(viii),\hfill}}$f(t)=\tfr12(t+1/t)$,
\hfill\break{\hbox to 30pt{(ix),\hfill}}$f(t)=\tfr14t(1+\tanh
t)$,\hfill\break
where $\hbox{Ai}(z)$ is the Airy function which is the solution of
$\hbox{Ai}''(z)-z\hbox{Ai}(z)=0$,
satisfying
$\hbox{Ai}(z)\sim \tfr12\pi^{-1/2}z^{-1/4}\exp\left(-\tfr23
z^{3/2}\right)$ as $z\to\infty$ and
$\hbox{Ai}(z)\sim \pi^{-1/2}|z|^{-1/4}\cos\left(\tfr23
|z|^{3/2}+\tfr14\pi\right)$ as $z\to-\infty$.

\smallskip\noindent{\bf Figure 2 ``Breather'' solutions.}
\hangindent=16pt\hangafter=1 The solution (\swwsoli) where
\hfill\break{\hbox to 30pt{(i),\hfill}}$f(t)=2\sin t + 10$,
\hfill\break{\hbox to 30pt{(ii),\hfill}}$f(t)=2\sin t + 1$,
\hfill\break{\hbox to 30pt{(iii),\hfill}}$f(t)=5\sin t + 3$,
\hfill\break{\hbox to 30pt{(iv),\hfill}}$f(t)=5\sin t + 1$.

\smallskip\noindent{\bf Figure 3.}
\hangindent=16pt\hangafter=1 (a) The solution (\swwsolii) with $c=2$
and (b), the solution
(\swwsmsol) with $\kappa=1.2$ and $a_0=a_1=a_2=1$.
\bye